\documentclass{article}

\title{A first look at transition amplitudes in $(2+1)$-dimensional causal dynamical triangulations}

\author{Joshua H. Cooperman \\ \emph{Department of Physics, University of California, Davis}\\ \\ Jonah M. Miller \\ \emph{Department of Physics, University of Colorado, Boulder}\\ }

\usepackage{float}
\usepackage{multicol}
\usepackage{subfigure}
\usepackage{tabularx}
\usepackage{booktabs}
\usepackage{graphicx}
\usepackage{amsmath}
\usepackage{latexsym}
\usepackage{mathrsfs}
\usepackage{geometry}
\usepackage{chngcntr}
\counterwithin{figure}{section}
\counterwithin{table}{section}
\numberwithin{equation}{section}

\begin{document}

\maketitle

\begin{abstract}
We study a lattice regularization of the gravitational path integral---causal dynamical triangulations---for $(2+1)$-dimensional Einstein gravity with positive cosmological constant in the presence of past and future spacelike boundaries of fixed intrinsic geometries. For spatial topology of a $2$-sphere, we determine the form of the Einstein-Hilbert action supplemented by the Gibbons-Hawking-York boundary terms within the Regge calculus of causal triangulations. Employing this action we numerically simulate a variety of transition amplitudes from the past boundary to the future boundary. 
To the extent that we have so far investigated them, these transition amplitudes 
appear consistent with the gravitational effective action previously found to characterize the ground state of quantum spacetime geometry 
within the Euclidean de Sitter-like phase. Certain of these transition amplitudes 
convincingly demonstrate that the so-called stalks present in this phase are numerical artifacts of the lattice regularization, 
seemingly indicate that the quantization technique of causal dynamical triangulations differs in detail from that of the no-boundary proposal of Hartle and Hawking, and 
possibly represent the first numerical simulations of portions of temporally unbounded quantum spacetime geometry within the causal dynamical triangulations approach. We also uncover tantalizing evidence suggesting that Lorentzian not Euclidean de Sitter spacetime dominates the ground state on sufficiently large scales.

\end{abstract}

\section{Introduction}\label{introduction}


A quantum theory's ground state dictates much but not all of that theory's structure: transition amplitudes above the ground state encode a rich structure themselves. Indeed, the information contained in transition amplitudes is most relevant for experimental tests of the quantum theory: for an experiment to have empirical import, it must strive to observe at least a modicum of change. As in all other quantum theories, one would like to compute transition amplitudes in a quantum theory of gravity. While such amplitudes have certainly been computed in various candidate quantum theories of gravity, we know of no such calculations within the causal dynamical triangulations approach, at least for spacetime dimension of $2+1$ or higher.\footnote{We note that Ambj\o rn \emph{et al} considered transition amplitudes within a spherical symmetry truncation of a variant of $(2+1)$-dimensional causal dynamical triangulations \cite{JA&JJ&RL&GV1,JA&JJ&RL&GV2} and that Benedetti \emph{et al} considered transition amplitudes in a more ordered variant of $(2+1)$-dimensional causal dynamical triangulations \cite{DB&RL&FZ}.} We remedy this situation by presenting here the first numerical simulations of transition amplitudes in causal dynamical triangulations.


Causal dynamical triangulations is a path integral based approach to the nonperturbative quantization of classical metric theories of gravity. (See \cite{JA&AG&JJ&RL3} for a review.) In rigorously defining this quantization, causal dynamical triangulations emulates the well established techniques of lattice quantum field theory, specifically, the introduction of a lattice regularization followed by application of finite size scaling and renormalization. The approach distinguishes itself in two related respects: its restriction on the spacetime geometries allowed to contribute to the gravitational path integral and its associated regularization of this path integral. One restricts to causal spacetime geometries, namely those admitting a global foliation by spacelike hypersurfaces all of the same topology. One regulates causal spacetime geometries by causal triangulations, namely Lorentzian simplicial manifolds possessing such a foliated structure. When computing the gravitational path integral according to this prescription, one thus additionally specifies a topology for the leaves of the foliation.

The restriction to causal spacetime geometries---the approach's key new feature---finds its primary motivation in the failures of previous (Euclidean) lattice quantum gravity programs to define physically sound quantum theories of gravity. (See, for instance, the review \cite{RL}.) With models in $1+1$ dimensions implicating the absence of any Lorentzian causal structure in these failures, Ambj\o rn, Loll, and Jurkiewicz decided to formulate a regularization of the gravitational path integral  directly in Lorentzian signature \cite{JA&JJ&RL1,JA&RL}. Of course, a generic Lorentzian spacetime geometry does not admit a global foliation by spacelike hypersurfaces all of the same topology, a fact that significantly complicates its regularization by a Lorentzian simplicial manifold. Consequently, these authors elected to impose the restriction to causal spacetime geometries. Besides allowing for the implementation of the lattice regularization in Lorentzian signature, this causality condition provides for a well defined Wick rotation to Euclidean signature required for Monte Carlo simulations, the primary means of investigation of causal dynamical triangulations. One may view the causality condition as a compromise between the necessities of introducing causal structure and of conducting numerical studies.

Causal dynamical triangulations has thus far produced several promising results. Most studies in $2+1$ and $3+1$ dimensions have focused on the ground state emerging from the quantization of  Einstein gravity with positive cosmological constant for topologically spherical leaves of the foliation.\footnote{There are two notable exceptions for $2+1$ dimensions: Budd considered topologically toric leaves \cite{TGB}, and Anderson \emph{et al} considered projectable Ho\v{r}ava-Lifshitz gravity \cite{CA&SJC&JHC&PH&RKK&PZ}.} 
Within the so-called phase C of quantum spacetime geometry, which is present for both $2+1$ and $3+1$ dimensions, this ground state exhibits the correct semiclassical behavior on sufficiently large scales and novel quantum mechanical behavior on sufficiently small scales. In particular, on large scales the gravitational effective action of a simple minisuperspace model describes exceedingly well the quantum spacetime geometry \cite{JA&JGS&AG&JJ,JA&AG&JJ&RL1,JA&AG&JJ&RL2,JA&AG&JJ&RL&JGS&TT,JA&JJ&RL3,JA&JJ&RL4,JA&JJ&RL5,JA&JJ&RL6}. The ensemble average spacetime geometry fits that of Euclidean de Sitter spacetime, the maximally symmetric extremum of this gravitational effective action \cite{JA&JGS&AG&JJ,JA&AG&JJ&RL1,JA&AG&JJ&RL2,JA&AG&JJ&RL&JGS&TT,JA&JJ&RL3,JA&JJ&RL4,JA&JJ&RL5,JA&JJ&RL6,CA&SJC&JHC&PH&RKK&PZ,DB&JH,RK}. Deviations from the ensemble average spacetime geometry fit a straightforward quantization of this gravitational effective action with Euclidean de Sitter spacetime as the ground state \cite{JA&AG&JJ&RL1,JA&AG&JJ&RL2}. 
Of course, these results accord with the semiclassical expectation that the ground state is the most symmetric configuration. On small scales the ensemble average spacetime geometry exhibits a dynamical dimensional reduction from the topological dimension of $2+1$ or $3+1$ to a dimension of approximately $2$ \cite{JA&JJ&RL6,JA&JJ&RL7,DB&JH,RK}. 
In $3+1$ dimensions there is also firm evidence for the presence of a second order phase transition at the B-C phase boundary, potentially pointing to a well defined continuum limit of causal dynamical triangulations \cite{JA&SJ&JJ&RL1,JA&SJ&JJ&RL2}.

All of these results concern the ground state of quantum spacetime geometry. We wish to learn more about the nature of the causal dynamical triangulations quantization scheme by studying transition amplitudes. Given the foliation of each causal triangulation into spacelike hypersurfaces of a fixed topology, one most readily considers transition amplitudes between two leaves of this foliation. Such amplitudes constitute the analogue in the quantum theory of the so-called thick sandwich problem: specify the intrinsic geometries of initial and final spacelike hypersurfaces, allowing their extrinsic geometries to vary, and then compute the evolution from initial to final spacelike hypersurface. In principle, one could also consider other types of transition amplitudes specified by other types of boundary conditions. For instance, Warner, Catterall, and Renken explored Euclidean dynamical triangulations with a single boundary, introducing a generalized boundary action with the hope of discovering an enriched phase structure \cite{SW&SC,SW&SC&RR}. 
We have not pursued these other possibilities for two primary reasons. First, their implementation in causal dynamical triangulations is rather more difficult than that of the thick sandwich problem. Second, the thick sandwich problem already affords the investigation of some interesting physical questions. Clearly, computing these transition amplitudes amounts to studying causal dynamical triangulations with fixed boundaries, a situation not previously explored for spacetime dimension of $2+1$ or higher.\footnote{Technically, we do not compute any transition amplitudes but only explore properties of representative causal triangulations contributing to certain transition amplitudes. The nature of Markov chain Monte Carlo simulations forces this approach on us.}

We specifically consider the causal dynamical triangulations of $(2+1)$-dimensional Einstein gravity with positive cosmological constant for spacetime topology of the direct product of a $2$-sphere $\mathcal{S}^{2}$ and a line interval $\mathcal{I}$. In section \ref{CDTwithboundaries} we first introduce the formalism of causal dynamical triangulations, and we then derive the form of the action appropriate to the boundary conditions under consideration. We also discuss the modified numerical implementation that we have devised to run simulations of the path integral with these boundary conditions.  In section \ref{transitionamplitudes} we report the results of our numerical simulations of three classes of transition amplitudes: from minimal initial boundary to minimal final boundary, from minimal initial boundary to nonminimal final boundary, and from nonminimal initial boundary to nonminimal final boundary. The second of these classes captures within causal dynamical triangulations the setting of the Hartle-Hawking no-boundary wavefunction \cite{JBH&SWH}. We demonstrate that the transition amplitudes of the first two classes and certain transition amplitudes of the third class are quantitatively consistent with the gravitational effective action previously found to characterize the ground state of quantum spacetime geometry within phase C. We propose that this may also be the case for all of the transition amplitudes of the third class. The transition amplitudes of the first class substantiate the hypothesis that the so-called stalks present in this phase are numerical artifacts of the lattice regularization. The transition amplitudes of the second class imply that the causal dynamical triangulations quantization scheme differs from that of the no-boundary proposal of Hartle and Hawking. Some of the transition amplitudes of the third class appear to constitute the first numerical simulations of portions of temporally unbounded quantum spacetime geometry within the causal dynamical triangulations approach. These last transition amplitudes also suggest that Lorentzian not Euclidean de Sitter spacetime may dominate the ground state of quantum spacetime geometry on sufficiently large scales. In section \ref{conclusion} we conclude by considering several extensions of our work and several questions that such extensions could address.

\section{Causal dynamical triangulations with fixed boundaries}\label{CDTwithboundaries}

\subsection{Formalism}

The path integral offers a powerful nonperturbative procedure for quantizing classical theories. To compute a transition amplitude $\mathcal{A}[\xi]$ in the quantum theory, one evaluates an expression of the form
\begin{equation}\label{pathintegral}
\mathcal{A}[\xi]=\int_{q|_{\mathcal{B}}=\xi}\mathcal{D}q\,e^{iS[q]}.
\end{equation}
Here, $q$ denotes the set of dynamical variables of the classical theory characterized by the action $S[q]$. One performs the path integration over all physically distinct configurations of the dynamical variables $q$ consistent with the boundary conditions $q|_{\mathcal{B}}=\xi$ defining the transition amplitude $\mathcal{A}[\xi]$. Of course, except in certain sufficiently simple circumstances, such path integrals are technically difficult not only to compute, but even to define.

\begin{figure}[!ht]
\centering
\includegraphics[scale=0.5]{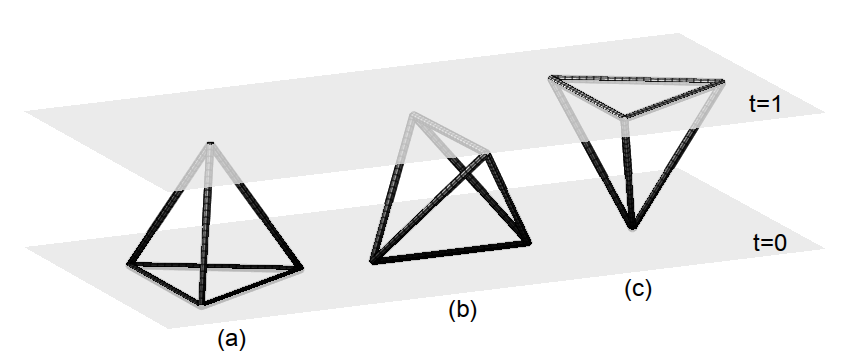}
\caption[Optional caption for list of figures]{The three types of $3$-simplices employed in $(2+1)$-dimensional causal dynamical triangulations: (a) $(3,1)$ $3$-simplex, (b) $(2,2)$ $3$-simplex, (c) $(1,3)$ $3$-simplex. The first number in the ordered pair indicates the number of vertices on the initial time slice ($t=0$), and the second number in the ordered pair indicates the number of vertices on the final time slice ($t=1$). We denote by $N_{3}^{(3,1)}$ the number of $(3,1)$ $3$-simplices, by $N_{3}^{(2,2)}$ the number of $(2,2)$ $3$-simplices, and by $N_{3}^{(1,3)}$ the number of $(1,3)$ $3$-simplices in a causal triangulation.}
\label{3-simplices}
\end{figure}

The causal dynamical triangulations approach nevertheless provides a prescription for overcoming these technical difficulties in the case of classical metric theories of gravity. A transition amplitude $\mathcal{A}[\mathbf{\gamma}]$ for such a theory takes the form
\begin{equation}\label{gravitypathintegral}
\mathcal{A}[\mathbf{\gamma}]=\int_{\mathbf{g}|_{\partial\mathcal{M}}=\mathbf{\gamma}}\mathcal{D}\mathbf{g}\,e^{iS[\mathbf{g}]},
\end{equation}
where $\mathbf{g}$ is the metric characterizing a spacetime $\mathcal{M}$ with boundary $\partial\mathcal{M}$. To define rigorously the transition amplitude \eqref{gravitypathintegral}, causal dynamical triangulations invokes a restriction on and a regularization of the spacetime geometries contributing to the path integral. Only spacetime geometries possessing the causal structure of a global foliation by spacelike hypersurfaces all of the same topology are permitted. One regulates these spacetime geometries by a specific class of simplicial manifolds, causal triangulations. A causal triangulation is a Lorentzian simplicial manifold constructed by gluing together Minkowskian simplices in a manner consistent with the presence of this foliation. We often refer to the foliation's leaves as time slices, and we enumerate them with a discrete Lorentzian time coordinate $t\in\{1,\ldots,T\}$. In $2+1$ dimensions, the case that we consider below, every causal triangulation is constructed from the set of Minkowskian $3$-simplices depicted in figure \ref{3-simplices}.
Each leaf of the foliation is triangulated by regular spacelike $2$-simplices---equilateral triangles---having squared edge length $a^{2}$. The fixed spacelike edge length $a$ serves as the lattice spacing. Adjacent leaves are then connected by timelike edges having squared length $-\alpha a^{2}$ for $\alpha>\frac{1}{2}$ such that only the three types of Minkowskian $3$-simplices are formed. Starting from a given causal triangulation $\mathcal{T}_{c}$ with a fixed number $T$ of time slices, one may obtain any other causal triangulation $\mathcal{T}'_{c}$ also with a fixed number $T$ of time slices by applying a finite sequence of the dynamical Pachner moves adapted to causal triangulations \cite{JA&JJ&RL1,JA&JJ&RL2}.\footnote{In fact, no one has proved that these Pachner moves are ergodic on the space of causal triangulations, but they are widely believed to possess this property.}

The prescription of causal dynamical triangulations transforms the path integral \eqref{gravitypathintegral} into the path sum
\begin{equation}\label{pathsum}
\mathcal{A}[\Gamma]=\sum_{\substack{\mathcal{T}_{c} \\ \mathcal{T}_{c}|_{\partial\mathcal{T}_{c}}=\Gamma}}\mu(\mathcal{T}_{c})e^{iS_{R}[\mathcal{T}_{c}]},
\end{equation}
where $\mu(\mathcal{T}_{c})$ is the measure, and $S_{R}[\mathcal{T}_{c}]$ is the action expressed in the Regge calculus of causal triangulations. One takes the measure equal to $\frac{1}{C(\mathcal{T}_{c})}$, the inverse of the order of the automorphism group of the causal triangulation $\mathcal{T}_{c}$. This factor accounts for the discrete remnants of the group of diffeomorphisms acting on the causal triangulation $\mathcal{T}_{c}$. The action $S_{R}[\mathcal{T}_{c}]$ depends on the classical theory of gravity that one wishes to quantize as well as the transition amplitude that one wishes to study, the latter dependence coming from the specification of boundary conditions by appropriate boundary terms in the action. 

\subsection{Action}

We are interested in the causal dynamical triangulations of $(2+1)$-dimensional Einstein gravity with positive cosmological constant. For spacetime manifolds $\mathcal{M}$ without boundary, this theory's action is just the Einstein-Hilbert action
\begin{equation}\label{EHaction}
S_{EH}[\mathbf{g}]=\frac{1}{16\pi G}\int_{\mathcal{M}}\mathrm{d}^{3}x\,\sqrt{-g}\left(R-2\Lambda\right).
\end{equation}
As observed by Gibbons and Hawking and by York, if the spacetime $\mathcal{M}$ has a boundary $\partial\mathcal{M}$, then one must supplement the action \eqref{EHaction} with a boundary term to ensure that variation of the total action leads to well posed equations of motion \cite{GWG&SWH,JWY}. The type of boundary conditions that one wants to enforce dictates the particular boundary term that one must include. If one wants to hold fixed the metric induced on the boundary $\partial\mathcal{M}$---the situation that we study below---then one must add to the action \eqref{EHaction} the Gibbons-Hawking-York boundary term
\begin{equation}
S_{GHY}[\mathbf{g}]=\frac{1}{8\pi G}\int_{\partial\mathcal{M}}\mathrm{d}^{2}y\sqrt{|\gamma|}K.
\end{equation}
Here, $\gamma$ is the determinant of the metric induced on the boundary $\partial\mathcal{M}$, $y$ denotes a set of coordinates on the boundary $\partial\mathcal{M}$, and $K$ is the trace of the extrinsic curvature of the boundary $\partial\mathcal{M}$. We thus employ the complete action
\begin{equation}\label{completeCaction}
S[\mathbf{g}]=S_{EH}[\mathbf{g}]+S_{GHY}[\mathbf{g}]
\end{equation}
in the path integral \eqref{gravitypathintegral}.

\begin{figure}[!ht]
\centering
\includegraphics[scale=0.5]{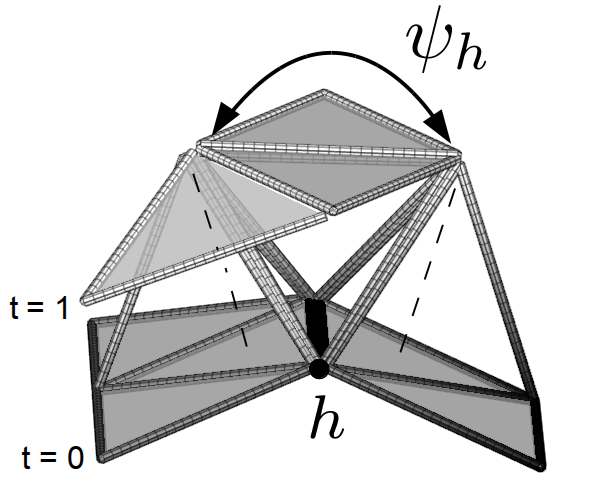}
\caption[Optional caption for list of figures]{An illustration of the construction of the extrinsic curvature for the hinge $h$ (black) indicating the angle $\psi_{h}$ through which the normal vector rotates for the case of $N_{3\uparrow}^{(2,2)}(h)=1$.}
\label{extrinsiccurvature}
\end{figure}

To apply the prescription of causal dynamical triangulations, we require the Regge calculus expression for the action \eqref{completeCaction}. Regge himself demonstrated that for a triangulation $\mathcal{T}$ the Einstein-Hilbert action \eqref{EHaction} assumes the form \cite{TR}
\begin{equation}\label{Reggeaction}
S_{EH}[\mathcal{T}]=\frac{1}{8\pi G}\sum_{h\in\mathcal{T}}V_{1}^{(h)}\delta_{h}-\frac{\Lambda}{8\pi G}\sum_{s\in\mathcal{T}}V_{3}^{(s)}.
\end{equation}
Here, $h$ is a hinge---a $1$-simplex in $2+1$ dimensions---of $1$-volume $V_{1}^{(h)}$ and deficit angle $\delta_{h}$, and $V_{3}^{(s)}$ is the spacetime $3$-volume of a $3$-simplex $s$. Hartle and Sorkin later determined the form of the Gibbons-Hawking-York boundary term in Regge calculus \cite{JBH&RS}. These authors showed that 
\begin{equation}\label{HSaction}
S_{GHY}[\mathcal{T}]=\frac{1}{8\pi G}\sum_{h\in\partial\mathcal{T}}V_{1}^{(h)}\psi_{h},
\end{equation}
where $h$ is hinge on the boundary $\partial\mathcal{T}$ of the triangulation $\mathcal{T}$ having $1$-volume $V_{1}^{(h)}$, and $\psi_{h}$ is the angle between the two vectors normal to the two spacelike $2$-simplices intersecting at the hinge $h$. See figure \ref{extrinsiccurvature}.

We now translate the expressions \eqref{Reggeaction} and \eqref{HSaction} into the Regge calculus of causal triangulations for $2$-sphere spatial topology and line interval temporal topology.\footnote{Ambj\o rn \emph{et al} performed this translation for the variant of $(2+1)$-dimensional causal dynamical triangulations considered in \cite{JA&JJ&RL&GV1,JA&JJ&RL&GV2}.} The reader not interested in the details of this construction may skip to equation \eqref{completeaction} in which we report our result. For an arbitrary causal triangulation $\mathcal{T}_{c}$ the Einstein-Hilbert action is
\begin{equation}
S_{EH}[\mathcal{T}_{c}]=\frac{1}{8\pi G}\sum_{h^{SL}\in\mathcal{T}_{c}}a\,\delta_{h^{SL}}+\frac{1}{8\pi G}\sum_{h^{TL}\in\mathcal{T}_{c}}\sqrt{\alpha}a\,\delta_{h^{TL}}-\frac{\Lambda}{8\pi G}\sum_{s\in\mathcal{T}_{c}}V_{3}^{(s)}
\end{equation}
since the $1$-volume $V_{1}^{(h^{SL})}$ of a spacelike hinge $h^{SL}$ is $a$, and the $1$-volume $V_{1}^{(h^{TL})}$ of a timelike hinge $h^{TL}$ is $\sqrt{\alpha}a$. Ambj\o rn \emph{et al} derived the specific form of the Einstein-Hilbert action for the case of spacetime topology $\mathcal{S}^{2}\times\mathcal{S}^{1}$, finding that
\begin{eqnarray}\label{periodicCDTaction}
S_{EH}[\mathcal{T}_{c}]&=&\frac{a}{8\pi G}\left[\frac{2\pi}{i}N_{1}^{SL}-\frac{2}{i}\theta_{SL}^{(2,2)}N_{3}^{(2,2)}-\frac{4}{i}\theta_{SL}^{(3,1)}N_{1}^{SL}+2\pi\sqrt{\alpha}N_{1}^{TL}-4\sqrt{\alpha}\theta_{TL}^{(2,2)}N_{3}^{(2,2)}\right.\nonumber\\ && \qquad\qquad\left.-3\sqrt{\alpha}\theta_{TL}^{(1,3)}N_{3}^{(1,3)}-3\sqrt{\alpha}\theta_{TL}^{(3,1)}N_{3}^{(3,1)}\right]\nonumber\\ && -\frac{\Lambda}{8\pi G}\left[V_{3}^{(2,2)}N_{3}^{(2,2)}+V_{3}^{(1,3)}N_{3}^{(1,3)}+V_{3}^{(3,1)}N_{3}^{(3,1)}\right].
\end{eqnarray}
Here, $N_{1}^{SL}$ is the number of spacelike $1$-simplices, $N_{1}^{TL}$ is the number of timelike $1$-simplices, $\theta_{SL}^{(p,q)}$ is the Lorentzian dihedral angle about a spacelike edge of a $(p,q)$ $3$-simplex, $\theta_{TL}^{(p,q)}$ is the Lorentzian dihedral angle about a timelike edge of a $(p,q)$ $3$-simplex, and $V_{3}^{(p,q)}$ is the Lorentzian spacetime volume of a $(p,q)$ $3$-simplex. 
We refer the reader to \cite{JA&JJ&RL2,CA&SJC&JHC&PH&RKK&PZ} for explicit expressions for these Lorentzian dihedral angles and spacetime volumes. Within the first set of square brackets, the first three terms stem from the summation over spacelike hinges, and the last four terms stem from the summation over timelike hinges. The terms within the second set of square brackets stem from the summation over $3$-simplices. As we demonstrate below, we must modify the action \eqref{periodicCDTaction} to account for the presence of the boundaries. 

We now derive the Gibbons-Hawking-York boundary term in this case. Given the spacetime topology $\mathcal{S}^{2}\times\mathcal{I}$, the boundary $\partial\mathcal{T}_{c}$ of a causal triangulation $\mathcal{T}_{c}$ consists of two disconnected components: an initial or past spatial $2$-sphere $\mathcal{S}_{i}^{2}$ and a final or future spatial $2$-sphere $\mathcal{S}_{f}^{2}$. Starting from the prescription \eqref{HSaction} of Hartle and Sorkin and drawing on a similar construction of Anderson \emph{et al} \cite{JBH&RS,CA&SJC&JHC&PH&RKK&PZ},\footnote{We subsequently learned that Dittrich and Loll had previously determined the relevant extrinsic curvature \cite{BD&RL}.} we propose the prescription
\begin{equation}\label{GHYCDTaction}
S_{GHY}[\mathcal{T}_{c}]=\frac{a}{8\pi G}\sum_{h\in\mathcal{S}_{i}^{2}}\frac{1}{i}\left[\pi-2\theta_{SL}^{(3,1)}-\theta_{SL}^{(2,2)}N_{3\uparrow}^{(2,2)}(h)\right]+\frac{a}{8\pi G}\sum_{h\in\mathcal{S}_{f}^{2}}\frac{1}{i}\left[\pi-2\theta_{SL}^{(1,3)}-\theta_{SL}^{(2,2)}N_{3\downarrow}^{(2,2)}(h)\right].
\end{equation}
Here, $N_{3\uparrow}^{(2,2)}(h)$ is the number of future-directed $(2,2)$ $3$-simplices attached to the hinge $h$, and $N_{3\downarrow}^{(2,2)}(h)$ is the number of past-directed $(2,2)$ $3$-simplices attached to the hinge $h$. We justify this prescription as follows. In parallel transporting the vector normal to one component of the boundary $\partial\mathcal{T}_{c}$ between two spacelike $2$-simplices intersecting at the hinge $h$, the vector rotates through the angle 
\begin{equation}
\frac{1}{i}\left[2\theta_{SL}^{(3,1)}+\theta_{SL}^{(2,2)}N_{3}^{(2,2)}(h)\right].
\end{equation}
See figure \ref{extrinsiccurvature}.
When this angle is $\frac{\pi}{i}$, the extrinsic curvature vanishes locally at the hinge $h$; this fact dictates the deficit angle-like form of our above prescription. One might have expected a relative negative sign in the Gibbons-Hawking-York boundary term \eqref{GHYCDTaction} between the contributions from the two disconnected components of the boundary. Its absence stems from the following fact: the future-directed orientation of the vector normal to $\mathcal{S}_{i}^{2}$ and the past-directed orientation of the vector normal to $\mathcal{S}_{f}^{2}$ are accounted for in the past-directed and future-directed orientations of the $(2,2)$ $3$-simplices attached to the boundaries. Performing the summations over the hinges on the boundary $\partial\mathcal{T}_{c}$, we may rewrite the Gibbons-Hawking-York boundary term as
\begin{eqnarray}
S_{GHY}[\mathcal{T}_{c}]&=&\frac{a}{8\pi G}\left[\frac{\pi}{i} N_{1}^{SL}(\mathcal{S}_{i}^{2})-\frac{2}{i}\theta_{SL}^{(3,1)}N_{1}^{SL}(\mathcal{S}_{i}^{2})-\frac{1}{i}\theta_{SL}^{(2,2)}N_{3\uparrow}^{(2,2)}(\mathcal{S}_{i}^{2})\right]\nonumber\\ &&+\frac{a}{8\pi G}\left[\frac{\pi}{i} N_{1}^{SL}(\mathcal{S}_{f}^{2})-\frac{2}{i}\theta_{SL}^{(1,3)}N_{1}^{SL}(\mathcal{S}_{f}^{2})-\frac{1}{i}\theta_{SL}^{(2,2)}N_{3\downarrow}^{(2,2)}(\mathcal{S}_{f}^{2})\right].
\end{eqnarray}

Before writing down the complete action for $(2+1)$-dimensional causal dynamical triangulations with $2$-sphere spatial topology and line interval temporal topology, we must account for the effect of the boundary on the Einstein-Hilbert term \eqref{periodicCDTaction}. In particular, the spacelike hinges on the boundary $\partial\mathcal{T}_{c}$ no longer contribute to the Einstein-Hilbert term because full spacetime parallel transport about those hinges is no longer well defined. Accordingly, we modify the terms stemming from the summation over spacelike hinges---the first three terms in the first set of square brackets in equation \eqref{periodicCDTaction}---of the above prescription for the Einstein-Hilbert term to
\begin{eqnarray}
&&\frac{a}{8\pi G}\left[\frac{2\pi}{i}\left(N_{1}^{SL}-N_{1}^{SL}(\mathcal{S}_{i}^{2})-N_{1}^{SL}(\mathcal{S}_{f}^{2})\right)\right.\nonumber\\ && \qquad\qquad \left.-\frac{1}{i}\theta_{SL}^{(2,2)}\left(2N_{3}^{(2,2)}-N_{3\uparrow}^{(2,2)}(\mathcal{S}_{i}^{2})-N_{3\downarrow}^{(2,2)}(\mathcal{S}_{f}^{2})\right)\right.\nonumber\\ && \qquad\qquad\qquad \left.-\frac{1}{i}\theta_{SL}^{(1,3)}\left(4N_{1}^{SL}-2N_{1}^{SL}(\mathcal{S}_{i}^{2})-2N_{1}^{SL}(\mathcal{S}_{f}^{2})\right)\right].
\end{eqnarray}
The numerical factors appearing in the second and third terms of this expression require some explanation. In the second term $N_{3\uparrow}^{(2,2)}(\mathcal{S}_{i}^{2})$ and $N_{3\downarrow}^{(2,2)}(\mathcal{S}_{f}^{2})$ enter with the factor $1$ instead of $2$ because only one of the two spacelike edges of these $(2,2)$ $3$-simplices attaches to the boundary $\partial\mathcal{T}_{c}$. In the third term $N_{1}^{SL}(\mathcal{S}_{i}^{2})$ and $N_{1}^{SL}(\mathcal{S}_{f}^{2})$ enter with the factor $2$ instead of $4$ because only two $(1,3)$ or $(3,1)$ $3$-simplices attach to a hinge on the boundary $\partial\mathcal{T}_{c}$. The complete action is thus
\begin{eqnarray}\label{completeaction}
S_{R}[\mathcal{T}_{c}]&=&\frac{a}{8\pi G}\left[\frac{2\pi}{i}\left(N_{1}^{SL}-N_{1}^{SL}(\mathcal{S}_{i}^{2})-N_{1}^{SL}(\mathcal{S}_{f}^{2})\right)-\frac{1}{i}\theta_{SL}^{(2,2)}\left(2N_{3}^{(2,2)}-N_{3\uparrow}^{(2,2)}(\mathcal{S}_{i}^{2})-N_{3\downarrow}^{(2,2)}(\mathcal{S}_{f}^{2})\right)\right.\nonumber\\ && \qquad\qquad\left.-\frac{1}{i}\theta_{SL}^{(1,3)}\left(4N_{1}^{SL}-2N_{1}^{SL}(\mathcal{S}_{i}^{2})-2N_{1}^{SL}(\mathcal{S}_{f}^{2})\right)+2\pi\sqrt{\alpha}N_{1}^{TL}-4\sqrt{\alpha}\theta_{TL}^{(2,2)}N_{3}^{(2,2)}\right.\nonumber\\ &&\left.\qquad\qquad-3\sqrt{\alpha}\theta_{TL}^{(1,3)}N_{3}^{(1,3)}-3\sqrt{\alpha}\theta_{TL}^{(3,1)}N_{3}^{(3,1)}\right]\nonumber\\ && -\frac{\Lambda}{8\pi G}\left[V_{3}^{(2,2)}N_{3}^{(2,2)}+V_{3}^{(1,3)}N_{3}^{(1,3)}+V_{3}^{(3,1)}N_{3}^{(3,1)}\right]\nonumber\\ && +\frac{a}{8\pi G}\left[\frac{\pi}{i} N_{1}^{SL}(\mathcal{S}_{i}^{2})-\frac{2}{i}\theta_{SL}^{(3,1)}N_{1}^{SL}(\mathcal{S}_{i}^{2})-\frac{1}{i}\theta_{SL}^{(2,2)}N_{3\uparrow}^{(2,2)}(\mathcal{S}_{i}^{2})\right]\nonumber\\ &&+\frac{a}{8\pi G}\left[\frac{\pi}{i} N_{1}^{SL}(\mathcal{S}_{f}^{2})-\frac{2}{i}\theta_{SL}^{(1,3)}N_{1}^{SL}(\mathcal{S}_{f}^{2})-\frac{1}{i}\theta_{SL}^{(2,2)}N_{3\downarrow}^{(2,2)}(\mathcal{S}_{f}^{2})\right].
\end{eqnarray}
Although at first glance the action \eqref{completeaction} does not appear to be real, it is in fact real for $\alpha>0$ given the expressions for the Lorentzian dihedral angles. In appendix \ref{composition} we demonstrate that the action \eqref{completeaction} is consistent with the composition of two causal triangulations sharing a common boundary. We use the action \eqref{completeaction} in the path sum \eqref{pathsum} to compute transition amplitudes $\mathcal{A}[\Gamma(\mathcal{S}_{i}^{2}),\Gamma(\mathcal{S}_{f}^{2})]$ between the initial and final spacelike boundaries $\mathcal{S}_{i}^{2}$ and $\mathcal{S}_{f}^{2}$ with fixed intrinsic geometries $\Gamma(\mathcal{S}_{i}^{2})$ and $\Gamma(\mathcal{S}_{f}^{2})$.

\subsection{Numerics}
\label{sec:numerical}

Although the formalism of causal dynamical triangulations has turned the computation of the path sum \eqref{pathsum} for the action \eqref{completeaction} into a well defined problem in combinatorics, solving this problem analytically even in $2+1$ dimensions is difficult. To study the transition amplitudes defined by the path sum \eqref{pathsum}, we thus turn to numerical methods, in particular, Markov chain Monte Carlo simulations of representative paths. To perform such numerical simulations, we must transform the path sum \eqref{pathsum} into a partition function with real as opposed to complex summands. As we noted above, the structure of causal triangulations is specifically adapted to such a transformation, allowing a simple Wick rotation from Lorentzian to Euclidean signature. This Wick rotation consists in analytically continuing the parameter $\alpha$ to $-\alpha$ in the lower half complex plane \cite{JA&JJ&RL2}. The path sum \eqref{pathsum} thus becomes the partition function
\begin{equation}\label{partitionfunction}
\mathcal{Z}[\Gamma]=\sum_{\substack{\mathcal{T}_{c} \\ \mathcal{T}_{c}|_{\partial\mathcal{T}_{c}}=\Gamma}}\mu(\mathcal{T}_{c})e^{-S_{R}^{(E)}[\mathcal{T}_{c}]}
\end{equation}
for the Euclidean action 
\begin{eqnarray}\label{completeEaction}
S_{R}^{(E)}[\mathcal{T}_{c}]&=&\frac{ia}{8\pi G}\left[\frac{2\pi}{i}\left(N_{1}^{SL}-N_{1}^{SL}(\mathcal{S}_{i}^{2})-N_{1}^{SL}(\mathcal{S}_{f}^{2})\right)-\frac{1}{i}\vartheta_{SL}^{(2,2)}\left(2N_{3}^{(2,2)}-N_{3\uparrow}^{(2,2)}(\mathcal{S}_{i}^{2})-N_{3\downarrow}^{(2,2)}(\mathcal{S}_{f}^{2})\right)\right.\nonumber\\ && \qquad\qquad\left.-\frac{1}{i}\vartheta_{SL}^{(1,3)}\left(4N_{1}^{SL}-2N_{1}^{SL}(\mathcal{S}_{i}^{2})-2N_{1}^{SL}(\mathcal{S}_{f}^{2})\right)-2\pi i\sqrt{-\alpha}N_{1}^{TL}+4i\sqrt{-\alpha}\vartheta_{TL}^{(2,2)}N_{3}^{(2,2)}\right.\nonumber\\ &&\left.\qquad\qquad+3i\sqrt{-\alpha}\vartheta_{TL}^{(1,3)}N_{3}^{(1,3)}+3i\sqrt{-\alpha}\vartheta_{TL}^{(3,1)}N_{3}^{(3,1)}\right]\nonumber\\ && -\frac{i\Lambda}{8\pi G}\left[\mathcal{V}_{3}^{(2,2)}N_{3}^{(2,2)}+\mathcal{V}_{3}^{(1,3)}N_{3}^{(1,3)}+\mathcal{V}_{3}^{(3,1)}N_{3}^{(3,1)}\right]\nonumber\\ && +\frac{ia}{8\pi G}\left[\frac{\pi}{i} N_{1}^{SL}(\mathcal{S}_{i}^{2})-\frac{2}{i}\vartheta_{SL}^{(3,1)}N_{1}^{SL}(\mathcal{S}_{i}^{2})-\frac{1}{i}\vartheta_{SL}^{(2,2)}N_{3\uparrow}^{(2,2)}(\mathcal{S}_{i}^{2})\right]\nonumber\\ &&+\frac{ia}{8\pi G}\left[\frac{\pi}{i} N_{1}^{SL}(\mathcal{S}_{f}^{2})-\frac{2}{i}\vartheta_{SL}^{(3,1)}N_{1}^{SL}(\mathcal{S}_{f}^{2})-\frac{1}{i}\vartheta_{SL}^{(2,2)}N_{3\downarrow}^{(2,2)}(\mathcal{S}_{f}^{2})\right],
\end{eqnarray}
Here, $\vartheta_{SL}^{(p,q)}$ is the Euclidean dihedral angle about a (still) spacelike edge of a $(p,q)$ $3$-simplex, $\vartheta_{TL}^{(p,q)}$ is the Euclidean dihedral angle about a (formerly) timelike edge of a $(p,q)$ $3$-simplex, and $\mathcal{V}_{3}^{(p,q)}$ is the Euclidean spacetime $3$-volume of a $(p,q)$ $3$-simplex. We refer the reader to \cite{JA&JJ&RL2,CA&SJC&JHC&PH&RKK&PZ} for explicit expressions for these Euclidean dihedral angles and spacetime volumes. When the initial and final spatial $2$-spheres $\mathcal{S}_{i}^{2}$ and $\mathcal{S}_{f}^{2}$ are identified, in which case the spacetime topology is $\mathcal{S}^{2}\times\mathcal{S}^{1}$, the action \eqref{completeEaction} assumes the simple form
\begin{equation}
S_{R}^{(E)}[\mathcal{T}_{c}]=-k_{0}N_{0}+k_{3}N_{3}
\end{equation}
for the couplings
\begin{subequations}\label{k0k3expressions}
\begin{eqnarray}
k_{0}&=&2\pi ak\\
k_{3}&=&\frac{a^{3}\lambda}{4\sqrt{2}}+2\pi ak\left[\frac{3}{\pi}\cos^{-1}{\left(\frac{1}{3}\right)}-1\right]
\end{eqnarray}
\end{subequations}
with $k=\frac{1}{8\pi G}$, $\lambda=\frac{\Lambda}{8\pi G}$, and $\alpha=-1$ \cite{JA&JJ&RL2,JA&JJ&RL3}. In $2+1$ dimensions the numerical value of $\alpha$ is irrelevant (so long as $|\alpha|>\frac{1}{2}$) because topological constraints render its value redundant \cite{JA&JJ&RL2,DB&JH}. We thus set $\alpha=-1$ in the following. When referring to an ensemble of causal triangulations with fixed initial and final boundaries $\mathcal{S}_{i}^{2}$ and $\mathcal{S}_{f}^{2}$, we employ the couplings $k_{0}$ and $k_{3}$ instead of the couplings $k$ and $\lambda$ to make contact with previous work. By the given values of $k_{0}$ and $k_{3}$, we mean the values dictated by the relations \eqref{k0k3expressions} for the values of $k$ and $\lambda$ actually characterizing the given ensemble. 

We perform numerical simulations as follows. The partition function \eqref{partitionfunction} involves a sum over all causal triangulations with the specified spacetime topology including those with arbitrarily large numbers $T$ of time slices and $N_{3}$ of $3$-simplices. For a given value of the coupling $k_{0}$, however, the partition function \eqref{partitionfunction} defines a critical surface for a critical value $k_{3}^{c}$ of the coupling $k_{3}$ on which the amplitude $\mu(\mathcal{T}_{c})e^{-S_{R}^{(E)}[\mathcal{T}_{c}]}$ is peaked for causal triangulations with particular fixed numbers $T$ of time slices and $N_{3}$ of $3$-simplices. Since we cannot simulate causal triangulations of arbitrarily large size, we work on a particular critical surface on which $T$ is precisely fixed and $N_{3}$ is approximately fixed. We thus numerically simulate the partition function
\begin{equation}\label{partitionfunctionfixedTN}
Z[\Gamma]=\sum_{\substack{\mathcal{T}_{c} \\ \mathcal{T}_{c}|_{\partial\mathcal{T}_{c}}=\Gamma \\ T(\mathcal{T}_{c})=T \\ N_{3}(\mathcal{T}_{c})=N_{3}}}\mu(\mathcal{T}_{c})e^{-S_{R}^{(E)}[\mathcal{T}_{c}]}
\end{equation}
for the action \eqref{completeEaction}. The partition function \eqref{partitionfunctionfixedTN} at fixed $T$ and $N_{3}$ is related by a Legendre transform to the partition function \eqref{partitionfunction} without these constraints \cite{JA&AG&JJ&RL3}:
\begin{equation}
\mathcal{Z}[\Gamma]=\sum_{\substack{T \\ N_{3}}}e^{-k_{3}N_{3}}Z[\Gamma].
\end{equation}
Following standard techniques, we generate multiple ensembles at different values of $T$ and $N_{3}$, exploring how physical observables finite size scale towards the continuum. Evidently, the dependence of $T$ and $N_{3}$ on $k_{0}$ is quite weak, so we can essentially select $T$ and $N_{3}$ independently of $k_{0}$ to within the numerical precision of $k_{0}$. Accordingly, we first select the number $T$ of time slices, the target number $\bar{N}_{3}$ of $3$-simplices, and the value of the coupling $k_{0}$.\footnote{One might be concerned that the transition amplitudes under consideration are not well defined with the number $T$ of time slices fixed. For the classical thick sandwich problem one does not specify any such property of the spacetime interpolating between the initial and final spacelike hypersurfaces. Indeed, the thick sandwich problem is generically not well posed if one specifies data beyond the intrinsic geometries of the initial and final spacelike hypersurfaces. We believe that this concern is not necessarily a problem within the quantum theory defined by causal dynamical triangulations. 
Previous results---specifically, the formation of so-called stalks---suggest that the temporal extent of the quantum spacetime geometry is determined dynamically.} 

We next construct the initial and final boundary $2$-spheres $\mathcal{S}_{i}^{2}$ and $\mathcal{S}_{f}^{2}$ to each possess the particular fixed intrinsic geometries $\Gamma(\mathcal{S}_{i}^{2})$ and $\Gamma(\mathcal{S}_{f}^{2})$. To incorporate the initial and final boundaries into a causal triangulation with $T$ time slices, we create a minimal causal triangulation over the time slices labelled by the Euclidean discrete time coordinate $\tau\in\{2,\ldots,T-1\}$. Within a minimal causal triangulation each time slice is a minimally triangulated $2$-sphere---the surface of a tetrahedron---and adjacent time slices are connected by the minimal number of $3$-simplices. We then employ an algorithm, described in appendix \ref{algorithm}, that connects the initial and final $2$-spheres, labelled by $\tau=1$ and $\tau=T$, to their adjacent time slices.\footnote{We gratefully acknowledge David Kamensky for developing this algorithm.} We finally apply the spacetime $3$-volume-increasing Pachner moves, always holding the boundary intrinsic geometries fixed, to raise the total number $N_{3}$ of $3$-simplices to the target value $\bar{N}_{3}$. Fortuitously, none of the Pachner moves can change the boundaries' intrinsic geometries.\footnote{Currently, we cannot design a boundary $2$-sphere with an arbitrarily chosen intrinsic geometry; nevertheless, the algorithm described in appendix \ref{algorithm} can take as input any boundary $2$-sphere.}

With the conditions for the simulation of a particular transition amplitude established, we now tune the coupling $k_{3}$ to its critical value $k_{3}^{c}$ so that we work on the critical surface on which the partition function \eqref{partitionfunctionfixedTN} is well defined. Portions of this critical surface may well fall into different phases of the partition function \eqref{partitionfunctionfixedTN}. Below we study exclusively phase C, the de Sitter-like phase of causal dynamical triangulations for $2$-sphere spatial topology. Floating point imprecision essentially guarantees that we cannot tune precisely to the critical surface, so we add to the action \eqref{completeEaction} a Lagrange multiplier term of the form $\epsilon |N_3-\bar{N}_3|$ for Lagrange multipier $\epsilon\ll 1$. Effectively, this term smears the critical surface into a thin critical volume. In the simulations reported below we have set $\epsilon=0.02$.

Once this initialization process is complete, we run a standard Metropolis algorithm based on the partition function \eqref{partitionfunctionfixedTN} for the action \eqref{completeEaction}. We employ the Pachner moves to update our simulations at each step, thereby moving through the space of causal triangulations with fixed boundaries, fixed $T$, and approximately fixed $N_{3}$. After a period of thermalization, we generate an ensemble of causal triangulations representative of their relative weightings in the partition function \eqref{partitionfunctionfixedTN} by sampling every five hundred sweeps, one sweep consisting of one attempted Pachner move for every $3$-simplex in the current causal triangulation. Once we have generated an ensemble, containing of order $10^{5}$ members in the results reported below, we estimate the expectation values of observables as ensemble averages.

Our computer code is a version of that reported in \cite{RK} modified to account for initial and final $2$-sphere boundaries with fixed intrinsic geometries. We can readily disable its fixed boundary functionality to allow for periodic boundary conditions, and we have verified that we recover results quantitatively in agreement with those reported previously. We believe that the results presented in the next section provide further evidence that our modified code is working properly. Although all of these results are for phase C, we have run our code with fixed boundaries for values of the coupling $k_{0}$ that should fall within phase A of the partition function \eqref{partitionfunctionfixedTN}, and we do indeed find that this is the case.

\section{Transition amplitudes}\label{transitionamplitudes}

We now report on our numerical simulation of a variety of transition amplitudes by the methods of causal dynamical triangulations. As discussed above, we exclusively consider the transition amplitudes $Z[\Gamma(\mathcal{S}_{i}^{2}),\Gamma(\mathcal{S}_{f}^{2})]$ from an initial spatial $2$-sphere $\mathcal{S}_{i}^{2}$ of fixed intrinsic geometry $\Gamma(\mathcal{S}_{i}^{2})$ to a final spatial $2$-sphere $\mathcal{S}_{f}^{2}$ of fixed intrinsic geometry $\Gamma(\mathcal{S}_{f}^{2})$, and we work at fixed numbers $T$ of time slices and $N_{3}$ of $3$-simplices. Specifically, we explore three classes of such transition amplitudes: from minimally triangulated $2$-sphere to minimally triangulated $2$-sphere, from minimally triangulated $2$-sphere to nonminimally triangulated $2$-sphere, and from nonminimally triangulated $2$-sphere to nonminimally triangulated $2$-sphere. Thus far we have limited our consideration to the dependence of the transition amplitudes $Z[\Gamma(\mathcal{S}_{i}^{2}),\Gamma(\mathcal{S}_{f}^{2})]$ on the discrete spatial $2$-volumes $N_{2}^{SL}(\mathcal{S}_{i}^{2})$ and $N_{2}^{SL}(\mathcal{S}_{f}^{2})$ of the initial and final boundaries as measured by the number $N_{2}^{SL}$ of spacelike $2$-simplices composing each triangulated $2$-sphere. That is, we have averaged the transition amplitudes $Z[\Gamma(\mathcal{S}_{i}^{2}),\Gamma(\mathcal{S}_{f}^{2})]$ over all of the geometrical degrees of freedom of $\mathcal{S}_{i}^{2}$ and $\mathcal{S}_{f}^{2}$ except for their discrete spatial $2$-volumes $N_{2}^{SL}(\mathcal{S}_{i}^{2})$ and $N_{2}^{SL}(\mathcal{S}_{f}^{2})$, obtaining the transition amplitudes $Z[N_{2}^{SL}(\mathcal{S}_{i}^{2}),N_{2}^{SL}(\mathcal{S}_{f}^{2})]$. 

We numerically simulate the geometry-averaged transition amplitudes $Z[N_{2}^{SL}(\mathcal{S}_{i}^{2}),N_{2}^{SL}(\mathcal{S}_{f}^{2})]$ as follows. We first employ a separate Markov chain Monte Carlo algorithm to generate a finite set of pairs of initial and final boundary $2$-spheres. Each $2$-sphere has the desired discrete spatial $2$-volume but otherwise random intrinsic geometry drawn from the uniform distribution.\footnote{Drawing the intrinsic geometries of the boundary $2$-spheres from the uniform distribution represents an assumption on our part, which we deemed acceptable for an initial investigation. We believe that the ensuing results---in particular, their agreement with previous results---justify our choice.} For each pair of initial and final boundary $2$-spheres, we then implement the numerical algorithm described above to generate an ensemble of representative causal triangulations. Finally, we compute ensemble average observables by averaging over the ensembles of representative causal triangulations for every pair of initial and final boundary $2$-spheres. In particular, we have employed sets consisting of ten pairs of initial and final boundary $2$-spheres. 

We have elected to study the transition amplitudes $Z[N_{2}^{SL}(\mathcal{S}_{i}^{2}),N_{2}^{SL}(\mathcal{S}_{f}^{2})]$ for two primary reasons. First of all, the numerical simulation of the full transition amplitudes $Z[\Gamma(\mathcal{S}_{i}^{2}),\Gamma(\mathcal{S}_{f}^{2})]$ necessitates methods for both characterization and generation of specific boundary geometries. We are in the process of developing such methods, but our numerical implementation does not currently possess these capabilities.
Moreover, most of the analytical computations of these transition amplitudes have been performed within a minisuperspace truncation of the metric degrees of freedom. Restricting consideration to the geometry-averaged transition amplitudes $Z[N_{2}^{SL}(\mathcal{S}_{i}^{2}),N_{2}^{SL}(\mathcal{S}_{f}^{2})]$ thus facilitates comparison of our results to those in the literature.

To study these transition amplitudes, we primarily perform numerical measurements of the observable $N_{2}^{SL}(\tau)$, the number $N_{2}^{SL}$ of spacelike $2$-simplices as a function of the discrete Euclidean time coordinate $\tau$. The value $N_{2}^{SL}$ for a particular value of $\tau$ is not an observable, even in the presence of fixed spacelike boundaries, since the value of $\tau$ is merely a labeling convention, typically differing from one representative causal triangulation to another. The set of values $N_{2}^{SL}(\tau)$ for all values of $\tau$ is an observable: only given the dynamics of $N_{2}^{SL}$ can one choose a consistent convention for the label $\tau$ across all representative causal triangulations.\footnote{There may be some gauge redundancy in the observable $N_{2}^{SL}(\tau)$, but this observable does contain physical information.} This observable essentially gives the discrete time evolution of the discrete spatial $2$-volume. 

By focusing on the observable $N_{2}^{SL}(\tau)$, we are anticipating that a minisuperspace truncation of the metric degrees of freedom will prove a good description of our results. As we remarked briefly above, there is now ample evidence that this truncation accurately describes the ground state geometry within phase C. In particular, the ensemble average spacetime geometry of this ground state on sufficiently large scales, characterized by the observable $N_{2}^{SL}(\tau)$, is remarkably well described by the gravitational effective action \cite{JA&JGS&AG&JJ,JA&AG&JJ&RL1,JA&AG&JJ&RL2,JA&AG&JJ&RL&JGS&TT,JA&JJ&RL3,JA&JJ&RL4,JA&JJ&RL5,JA&JJ&RL6}
\begin{equation}\label{effectiveaction}
S_{\mathrm{eff}}^{(E)}[N_{2}^{SL}(\tau)]=c_{1}\sum_{\tau=1}^{T}\left[\frac{1}{N_{2}^{SL}(\tau)}\left(\frac{\Delta N_{2}^{SL}(\tau)}{\Delta\tau}\right)^{2}+c_{2}\right].
\end{equation}
Here, $c_{1}$ and $c_{2}$ are phenomenological couplings, and $\frac{\Delta}{\Delta\tau}$ represents an appropriate discrete time derivative. The effective action \eqref{effectiveaction} is the discrete analogue of the Einstein-Hilbert action
\begin{equation}\label{minisuperspaceaction}
S_{EH}^{(E)}[\rho(\eta)]=\frac{1}{2G}\int\mathrm{d}\eta\sqrt{g_{\eta\eta}}\left[g^{\eta\eta}\left(\frac{\mathrm{d}\rho(\eta)}{\mathrm{d}\eta}\right)^{2}+1-\Lambda\rho^{2}(\eta)\right]
\end{equation}
for a Euclidean minisuperspace model described by the line element
\begin{equation}\label{minisuperspacemetric}
\mathrm{d}s^{2}=g_{\eta\eta}\mathrm{d}\eta^{2}+\rho^{2}(\eta)\left(\mathrm{d}\theta^{2}+\sin^{2}{\theta}\mathrm{d}\phi^{2}\right)
\end{equation}
for scale factor $\rho$ as a function of the global time coordinate $\eta$. The discrete analogue of the Lagrange multiplier term $\Lambda \rho^{2}(\eta)$ in the action \eqref{minisuperspaceaction} does not appear in the effective action \eqref{effectiveaction} because each causal triangulation within a given ensemble is characterized by a fixed number $\bar{N}_{3}$ of $3$-simplices. We retain the constant factor $g_{\eta\eta}$ for convenience in making a correspondence with the discrete effective action \eqref{effectiveaction}. The maximally symmetric spacetime characterized by the line element \eqref{minisuperspacemetric} is Euclidean de Sitter spacetime for which
\begin{equation}\label{dSscalefactor}
\rho(\eta)=l_{dS}\cos{\left(\frac{\sqrt{g_{\eta\eta}}\eta}{l_{dS}}\right)}
\end{equation}
for the de Sitter radius $l_{dS}=\sqrt{\frac{1}{\Lambda}}$. The solution \eqref{dSscalefactor} describes a $3$-sphere of radius $l_{dS}$. The ensemble average $\langle N_{2}^{SL}(\tau)\rangle$ is exceedingly well fit by the discrete analogue of the time evolution of the spatial $2$-volume $V_{2}(\eta)=4\pi\rho^{2}(\eta)$ of Euclidean de Sitter spacetime \cite{JA&JGS&AG&JJ,JA&AG&JJ&RL1,JA&AG&JJ&RL2,JA&AG&JJ&RL&JGS&TT,JA&JJ&RL3,JA&JJ&RL4,JA&JJ&RL5,JA&JJ&RL6,CA&SJC&JHC&PH&RKK&PZ,DB&JH,RK}, explicitly
\begin{equation}\label{discretedSvolprofile}
\mathscr{N}_{2}^{SL}(\tau)=\frac{2}{\pi}\frac{\langle N_{3}^{(1,3)}\rangle}{\tilde{s}_{0}\langle N_{3}^{(1,3)}\rangle^{1/3}}\cos^{2}{\left(\frac{\tau}{\tilde{s}_{0}\langle N_{3}^{(1,3)}\rangle^{1/3}}\right)}.
\end{equation}
We demonstrate the explicit correspondence between the function \eqref{discretedSvolprofile} and the scale factor of Euclidean de Sitter spacetime in appendix \ref{volumeprofilederivation}. The parameter $\tilde{s}_{0}$ is fit to the particular ensemble of causal triangulations, serving to relate the dimensionless de Sitter radius to the scale of the discrete time coordinate $\tau$. 

Assuming for the moment that this minisuperspace truncation similarly applies to the transition amplitudes treated below, we consider briefly the semiclassical expectations for these transition amplitudes. 
Several authors have argued that one can compute the Lorentzian path integral \eqref{gravitypathintegral} by performing an appropriate Euclidean path integral. (See, for instance, \cite{JBH&SWH}.) When considering the Euclidean path integral within the minisuperspace truncation, one typically calculates the semiclassical approximation to this path integral by the method of steepest descents. For the transition amplitudes of interest, this path integral is 
\begin{equation}\label{pathintegralscalefactor}
\mathcal{A}[\rho_{i},\rho_{f}]=\int_{\rho(\eta_{i})=\rho_{i}}^{\rho(\eta_{f})=\rho_{f}}\mathcal{D}\rho(\eta)\,e^{-S_{EH}^{(E)}[\rho(\eta)]}.
\end{equation}
We consider the semiclassical predictions of the path integral \eqref{pathintegralscalefactor} for each of the three classes of transition amplitudes explored below.

\begin{figure}[!ht]
\centering
\subfigure[ ]{
\includegraphics[scale=0.25]{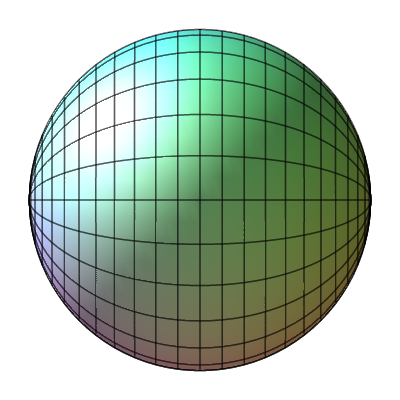}
\label{completesphere}
}
\subfigure[ ]{
\includegraphics[scale=0.225]{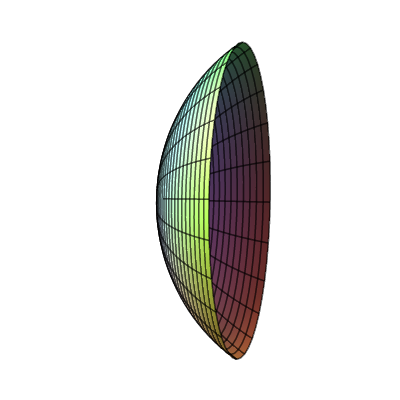}
\label{lessthansphere}
}
\subfigure[ ]{
\includegraphics[scale=0.31]{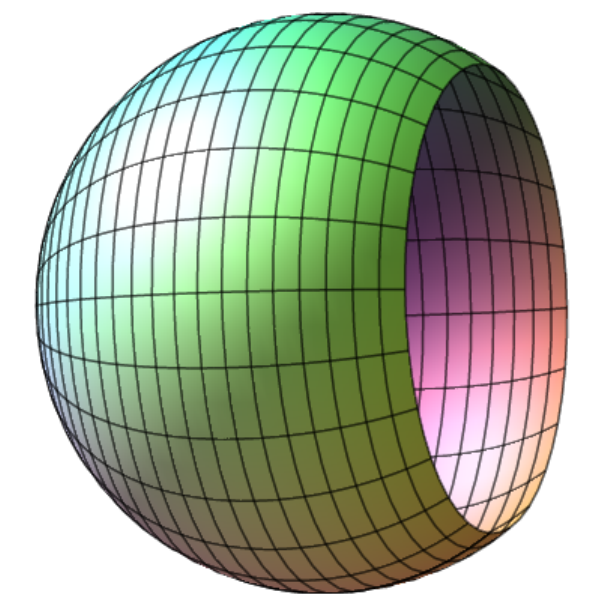}
\label{morethansphere}
}
\subfigure[ ]{
\includegraphics[scale=0.325]{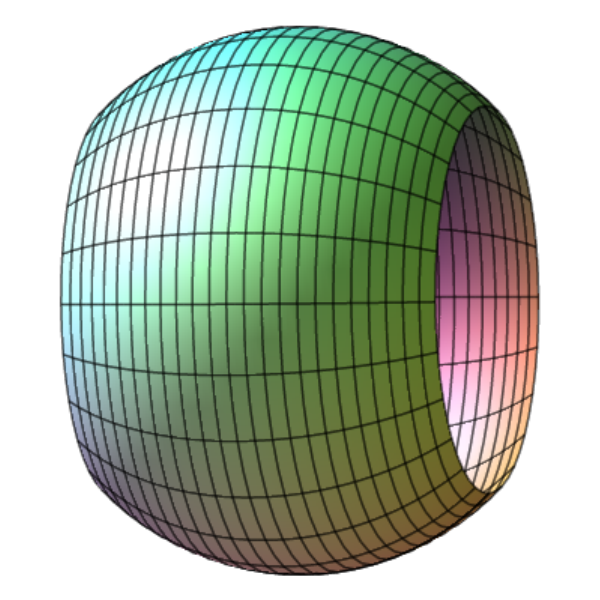}
\label{portionofsphere}
}
\caption[Optional caption for list of figures]{Extrema of the action \eqref{minisuperspaceaction} for $\rho_{i}<l_{dS}$ and $\rho_{f}<l_{dS}$ depicted for one fewer spacelike dimension. \subref{completesphere} $\rho_{i}=0$, $\rho_{f}=0$ \subref{lessthansphere} $\rho_{i}=0$, $\rho_{f}\neq0$ \subref{morethansphere} $\rho_{i}=0$, $\rho_{f}\neq0$ \subref{portionofsphere} $\rho_{i}\neq0$, $\rho_{f}\neq0$}
\label{extrema}
\end{figure}

For the transition amplitudes from minimally triangulated $2$-sphere to minimally triangulated $2$-sphere, corresponding to $\rho_{i}=0$ and $\rho_{f}=0$, Euclidean de Sitter spacetime is the extremum of the action \eqref{minisuperspaceaction} that dominates the steepest descents approximation of the path integral \eqref{pathintegralscalefactor}. See figure \ref{extrema}\subref{completesphere}. This accords of course with previous results for the ground state within phase C. For the transition amplitudes from minimally triangulated $2$-sphere to nonminimally triangulated $2$-sphere, corresponding to the Hartle-Hawking no-boundary scenario of $\rho_{i}=0$, the dominant extrema of the action \eqref{minisuperspaceaction} depend on the relative values of $\rho_{f}$ and $l_{dS}$. If $\rho_{f}<l_{dS}$, then there are two real extrema of the action \eqref{minisuperspaceaction} corresponding to the two portions of Euclidean de Sitter spacetime with final scale factor $\rho_{f}$, one containing less and one containing more than half of the spacetime $3$-volume of Euclidean de Sitter spacetime. See figures \ref{extrema}\subref{lessthansphere} and \ref{extrema}\subref{morethansphere}.
If $\rho_{f}=l_{dS}$, then these two extrema coincide. If $\rho_{f}>l_{dS}$, then there are two dominant complex extrema of the action \eqref{minisuperspaceaction}, yielding an oscillatory transition amplitude. The envelope of these oscillations gives a probability distribution for $\rho_{f}$ consistent with that of Lorentzian de Sitter spacetime. For the transition amplitude from nonminimally triangulated $2$-sphere to nonminimally triangulated $2$-sphere, the dominant extrema of the action depend on the relative values of $\rho_{i}$, $\rho_{f}$, and $\l_{dS}$. In particular, if $\rho_{i}<l_{dS}$ and $\rho_{f}<l_{dS}$, then the dominant extremum of the action \eqref{minisuperspaceaction} corresponds to a portion of Euclidean de Sitter spacetime with initial scale factor $\rho_{i}$ and final scale factor $\rho_{f}$. See figure \ref{extrema}\subref{portionofsphere}. If $\rho_{i}>l_{dS}$ and $\rho_{f}>l_{dS}$, then the dominant extrema of the action \eqref{minisuperspaceaction} are complex, again yielding an oscillatory transition amplitude. The envelope of these oscillations gives a probability distribution for $\rho_{i}$ and $\rho_{f}$ consistent with that of Lorentzian de Sitter spacetime.  

Given that the Markov chain Monte Carlo method yields an ensemble of representative paths contributing to the path integral and that the semiclassical contributions dominate the path integral, we expect our numerical simulations to output paths close to the semiclassical expectation. In analyzing transition amplitudes within causal dynamical triangulations, we attempt to determine if this is the case. 
 


\subsection{Minimal initial and final boundaries}

We first take both the initial boundary $\mathcal{S}_{i}^{2}$ and the final boundary $\mathcal{S}_{f}^{2}$ as the minimal triangulation of the $2$-sphere, namely the surface of the regular tetrahedron. (Recall that the causality condition on causal triangulations dictates that every time slice has the topology of a $2$-sphere.) The minimal triangulation of the $2$-sphere represents the best approximation to a no-boundary boundary condition within causal dynamical triangulations. For ensembles characterized by a sufficiently large number $\bar{N}_{3}$ of $3$-simplices, the nonzero spatial extent of the minimal boundary is presumably negligible, as our numerical simulations attest.

\begin{figure}[!ht]
\centering
\subfigure[ ]{
\includegraphics[scale=0.405]{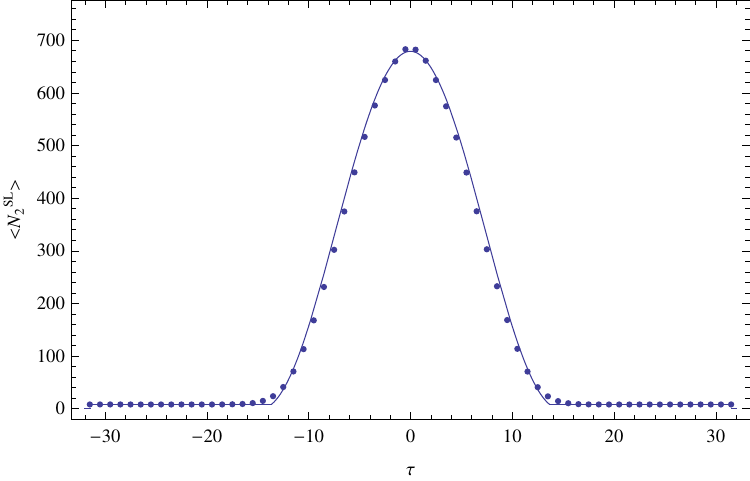}
\label{periodicdSvolfit64slices}
}
\subfigure[ ]{
\includegraphics[scale=0.405]{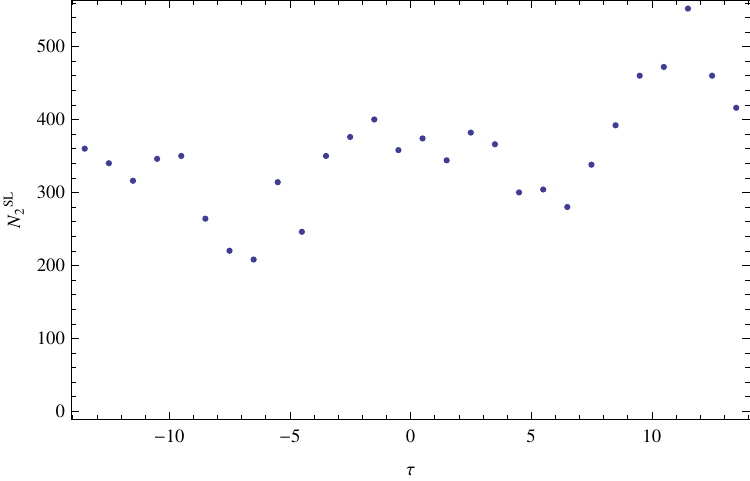}
\label{periodicvol28slices}
}
\caption[Optional caption for list of figures]{\subref{dSvolfit64slices} Coherent ensemble average number $\langle N_{2}^{SL}\rangle$ of spacelike $2$-simplices as a function of the discrete time coordinate $\tau$ for $T=64$, $\bar{N}_{3}=30850$, and $k_{0}=1.00$. \subref{periodicvol28slices} Number $N_{2}^{SL}$ of spacelike $2$-simplices as a function of the discrete time coordinate $\tau$ for a representative causal triangulation with $T=28$, $\bar{N}_{3}=30850$, and $k_{0}=1.00$.}
\label{comparisons}
\end{figure}

\begin{figure}[!ht]
\centering
\subfigure[ ]{
\includegraphics[scale=0.405]{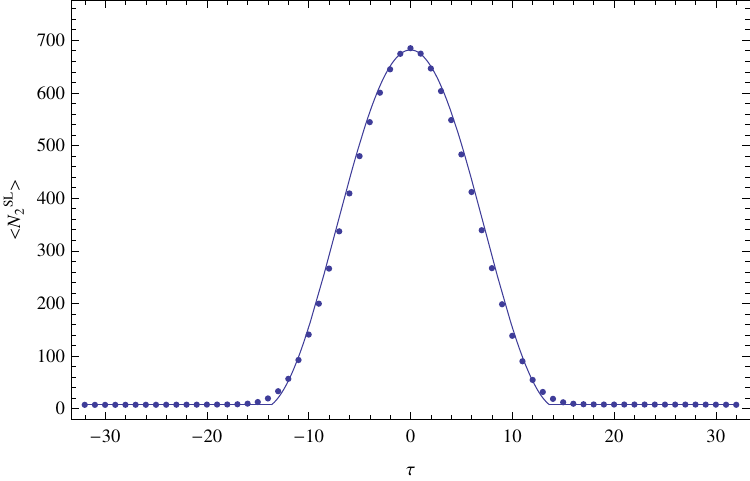}
\label{dSvolfit64slices}
}
\subfigure[ ]{
\includegraphics[scale=0.405]{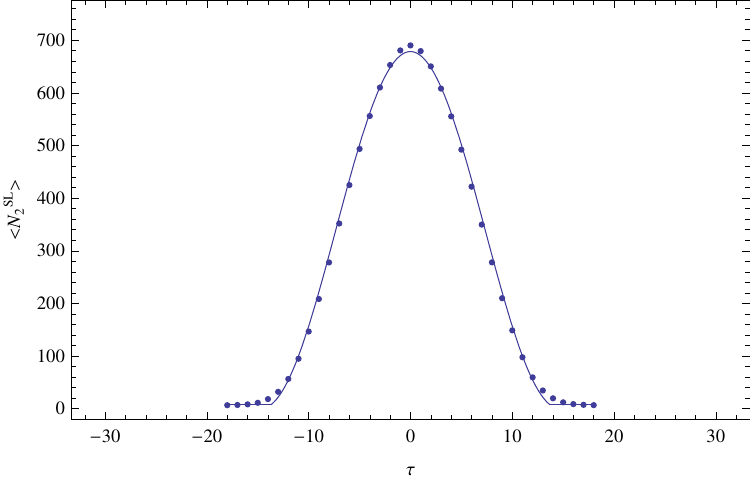}
\label{dSvolfit36slices}
}
\subfigure[ ]{
\includegraphics[scale=0.405]{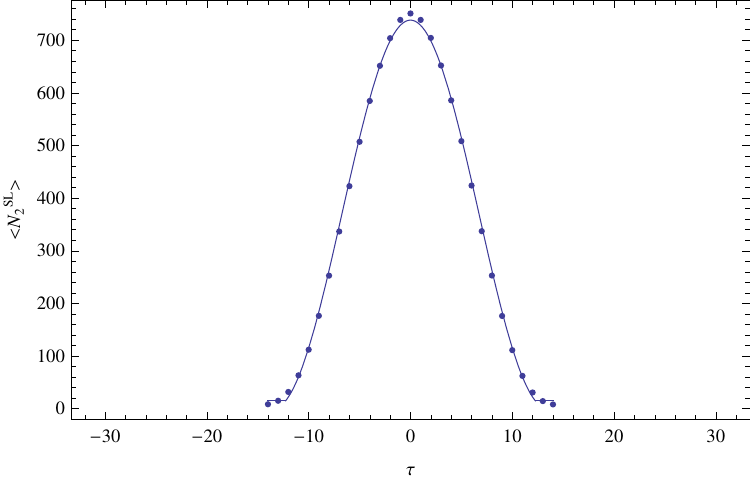}
\label{dSvolfit28slices}
}
\caption[Optional caption for list of figures]{Coherent ensemble average number $\langle N_{2}^{SL}\rangle$ of spacelike $2$-simplices as a function of the discrete time coordinate $\tau$ for $\bar{N}_{3}=30850$ and $k_{0}=1.00$. \subref{dSvolfit64slices} $T=65$ \subref{dSvolfit36slices} $T=37$  \subref{dSvolfit28slices} $T=29$}
\label{minminvaryingTstalks}
\end{figure}

Before displaying the results of our simulations, we recall the findings of two simulations with periodic boundary conditions for the Euclidean discrete time coordinate $\tau$. One may regard these as simulations of vacuum persistence amplitudes as opposed to transition amplitudes. In figure \ref{comparisons}\subref{dSvolfit64slices} we display $\langle N_{2}^{SL}(\tau)\rangle$, the coherent ensemble average number $\langle N_{2}^{SL}\rangle$ of spacelike $2$-simplices as a function of $\tau$, for an ensemble characterized by $T=64$, $\bar{N}_{3}=30850$, and $k_{0}=1.00$. Points indicate numerical measurements of $\langle N_{2}^{SL}\rangle$; error bars indicating the standard deviation $\sigma(\langle N_{2}^{SL}\rangle)$ are not visible on the scale of the plot. Clearly, $\langle N_{2}^{SL}(\tau)\rangle$ possesses a distinctly regular form. There is a central accumulation of discrete spatial $2$-volume spanning a significant portion of the time slices. The time slices beyond this accumulation constitute the so-called stalk in which each time slice has a near minimal number of spacelike $2$-simplices. In figure \ref{comparisons}\subref{dSvolfit64slices}, for instance, the central accumulation spans the time slices for which $\tau\in\{-14,\ldots,14\}$, and the stalk spans the time slices for which $\tau\in\{-32,\ldots,-15\}\cup\{15,\ldots,32\}$. This form of $\langle N_{2}^{SL}(\tau)\rangle$ is the characteristic feature of the de Sitter-like phase C. The curve in figure \ref{comparisons}\subref{dSvolfit64slices} is a fit of the function \eqref{discretedSvolprofile} to $\langle N_{2}^{SL}(\tau)\rangle$ within the central accumulation of discrete spatial $2$-volume. Within the stalk we fit a constant function to $\langle N_{2}^{SL}(\tau)\rangle$, matching appropriately at the junctions with the central accumulation. In appendix \ref{volumeprofilederivation} we explain the details of this and related fits.
The stalk is believed to be an artifact of the causality condition; interestingly, however, the effective action \eqref{effectiveaction} evidently provides a good description of the stalk too \cite{JA&JGS&AG&JJ,JA&AG&JJ&RL&JGS&TT}.\footnote{When the Euclidean discrete time coordinate $\tau$ is periodically identified, the center of discrete spatial $2$-volume of the central accumulation may occur at any value of $\tau$. Accordingly, to coherently average $N_{2}^{SL}(\tau)$ over an ensemble of causal triangulations, we temporally align each causal triangulation's center of discrete spatial $2$-volume \cite{CA&SJC&JHC&PH&RKK&PZ}.}

In figure \ref{comparisons}\subref{periodicvol28slices} we display $N_{2}^{SL}(\tau)$ for a single representative causal triangulation from an ensemble characterized by $T=28$, $\bar{N}_{3}=30850$, and $k_{0}=1.00$. Note that the central accumulation of discrete $2$-volume in figure \ref{comparisons}\subref{dSvolfit64slices} extends over approximately twenty-eight time slices. In this case there are evidently too few time slices for the de Sitter-like phase to form. Instead, representative causal triangulations have an approximately uniform distribution of discrete spatial $2$-volume over all of the time slices. 

We now consider a set of ensembles for which both the initial boundary $\mathcal{S}_{i}^{2}$ and the final boundary $\mathcal{S}_{f}^{2}$ are fixed to the minimal triangulation of the $2$-sphere. In particular, we set $\bar{N}_{3}=30850$ and $k_{0}=1.00$ for successively fewer numbers $T$ of time slices. In figure \ref{minminvaryingTstalks} we depict $\langle N_{2}^{SL}(\tau)\rangle$ for three such ensembles with $T=65$, $T=37$, and $T=29$ fit to the function \eqref{discretedSvolprofile}. For these numbers $T$ of time slices, representative causal triangulations still possess a stalk---technically, two disconnected stalks, one on either side of the central accumulation---although its extent diminishes as $T$ decreases. We also note that we must still coherently average $N_{2}^{SL}(\tau)$ despite the fixed boundaries, which do not temporally align the centers of discrete spatial $2$-volume in these cases. 

\begin{figure}[!ht]
\centering
\subfigure[ ]{
\includegraphics[scale=0.405]{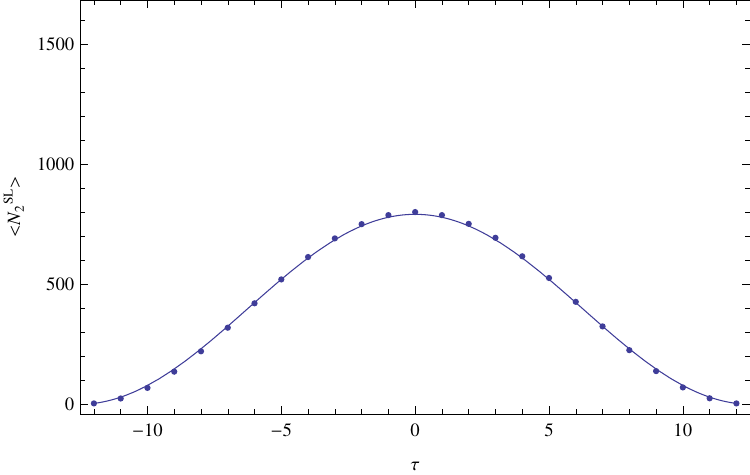}
\label{dSvolfit24slices}
}
\subfigure[ ]{
\includegraphics[scale=0.405]{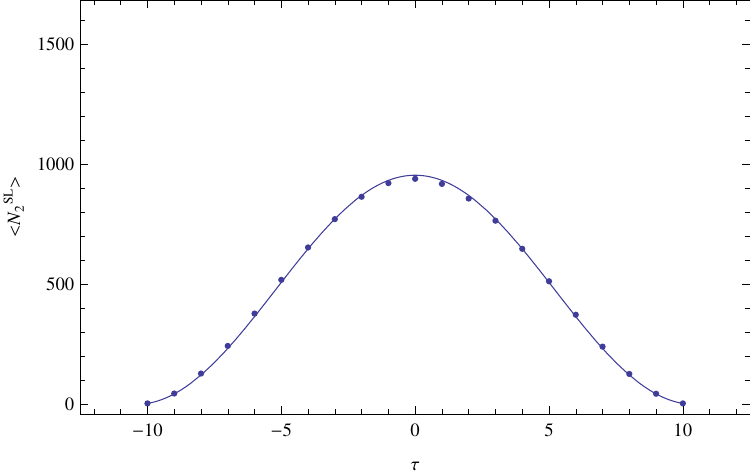}
\label{dSvolfit20slices}
}\\
\subfigure[ ]{
\includegraphics[scale=0.405]{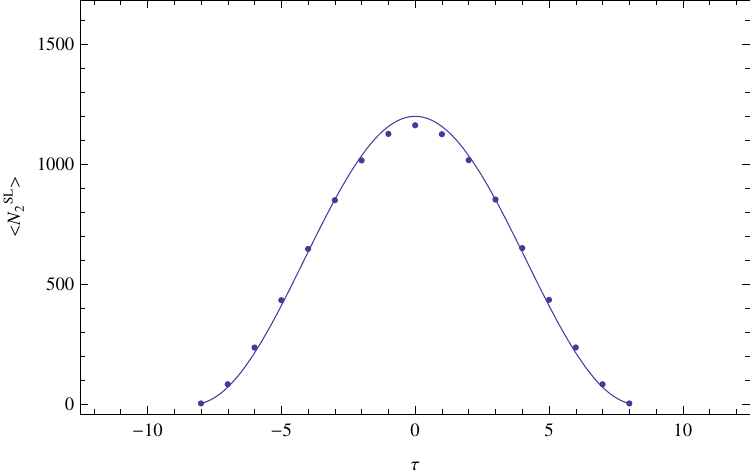}
\label{dSvolfit16slices}
}
\subfigure[ ]{
\includegraphics[scale=0.405]{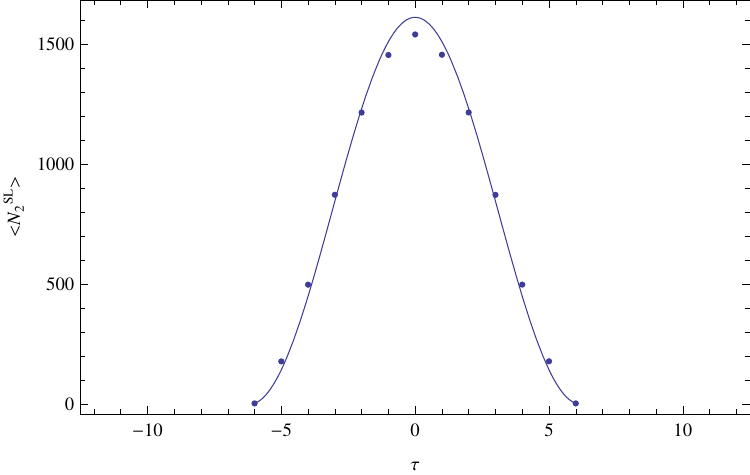}
\label{dSvolfit12slices}
}
\caption[Optional caption for list of figures]{Ensemble average number $\langle N_{2}^{SL}\rangle$ of spacelike $2$-simplices as a function of the discrete time coordinate $\tau$ for $\bar{N}_{3}=30850$ and $k_{0}=1.00$. \subref{dSvolfit24slices} $T=25$ \subref{dSvolfit20slices} $T=21$ \subref{dSvolfit16slices} $T=17$ \subref{dSvolfit12slices} $T=13$}
\label{minminvaryingT}
\end{figure}

In figure \ref{minminvaryingT} we display $\langle N_{2}^{SL}(\tau)\rangle$ for four more such ensembles with $T=25$, $T=21$, $T=17$, and $T=13$ fit to the function \eqref{discretedSvolprofile}. For these numbers $T$ of time slices, representative causal triangulations no longer possess stalks, and the fixed boundaries serve  to temporally align the centers of discrete spatial $2$-volume, effectively performing the coherent averaging automatically. One can thus completely remove the stalk, demonstrating that it is indeed an artifact of the numerical implementation of the causality condition. Moreover, as one further reduces the number of time slices, $\langle N_{2}^{SL}(\tau)\rangle$ dynamically readjusts to fit the function \eqref{discretedSvolprofile} for different values of $\tilde{s}_{0}$. 

Visually, all of the fits of the function \eqref{discretedSvolprofile} to $\langle N_{2}^{SL}(\tau)\rangle$ appear quite good, but we wish to ascertain the quality of these fits quantitatively. To this end we have determined the goodness of fit, as measured by the chi-squared per degree of freedom $\chi_{\mathrm{pdf}}^{2}$, for a variety of ensembles. In table \ref{goodnessoffit} we list the value of $\chi_{\mathrm{pdf}}^{2}$ for each of these ensembles. In appendix \ref{volumeprofilederivation} we explain the details of this statistical analysis.
\begin{table}[!ht]
\centering
\begin{tabularx}{13.75 cm}{cccccccc}
\toprule
Topology & $\bar{N}_{3}$ & $T$ & $N(\mathcal{T}_{c})$ & $\Delta\tau_{\mathrm{stalk}}/T$ & $\tilde{s}_{0}\pm\epsilon({\tilde{s}_{0}})$& $\chi_{\mathrm{pdf}}^{2}$ & $\chi_{\mathrm{pdf}}^{2}/N(\mathcal{T}_{c})$ \\
\toprule
$\mathcal{S}^{2}\times\mathcal{S}^{1}$ & $30850$ & $64$ & $1\cdot10^{4}$ & $0.5663$ & $0.4343$ &  $539.71$ & $0.0539$ \\
$\mathcal{S}^{2}\times\mathcal{S}^{1}$ & $65540$ & $96$ & $1\cdot10^{3}$ & $0.5988$ & $0.4649$ &  $92.74$ & $0.0927$ \\
$\mathcal{S}^{2}\times\mathcal{S}^{1}$ & $102400$ & $144$ & $3\cdot10^{3}$ & $0.7033$ & $0.4414$ & $162.89$ & $0.0524$ \\
\midrule
$\mathcal{S}^{2}\times\mathcal{I}$ & $30850$ & $13$ & $1\cdot10^{4}$ & $0$ & $0.1831$ &  $2591.72$ & $0.2590$ \\
$\mathcal{S}^{2}\times\mathcal{I}$ & $30850$ & $17$ & $1\cdot10^{4}$ & $0$ & $0.2457$ & $691.37$ & $0.0690$ \\
$\mathcal{S}^{2}\times\mathcal{I}$ & $30850$ & $21$ & $1\cdot10^{4}$ & $0$ & $0.309^{+0.004}$ &  $79.91$ & $0.0008$ \\
$\mathcal{S}^{2}\times\mathcal{I}$ & $30850$ & $25$ & $1\cdot10^{4}$ & $0$ & $0.37_{-0.01}$ &  $152.52$ & $0.0152$ \\
$\mathcal{S}^{2}\times\mathcal{I}$ & $30850$ & $29$ & $1\cdot10^{4}$ & $0.1230$ & $0.3992$ &  $752.65$ & $0.0751$ \\
$\mathcal{S}^{2}\times\mathcal{I}$ & $30850$ & $37$ & $1\cdot10^{4}$ & $0.2416$ & $0.4344$ &  $1901.58$ & $0.1612$ \\
$\mathcal{S}^{2}\times\mathcal{I}$ & $30850$ & $65$ & $1\cdot10^{4}$ & $0.5739$ & $0.4326$ &  $1157.59$ & $0.1156$ \\
$\mathcal{S}^{2}\times\mathcal{I}$ & $65540$ & $30$ & $1\cdot10^{3}$ & $0$ & $0.358_{-0.002}$ &  $16.95$ & $0.0168$ \\
$\mathcal{S}^{2}\times\mathcal{I}$ & $65540$ & $35$ & $5\cdot10^{3}$ & $0.1808$ & $0.388_{-0.001}^{+0.001}$ & $216.75$ & $0.0427$ \\
$\mathcal{S}^{2}\times\mathcal{I}$ & $102400$ & $32$ & $1\cdot10^{3}$ & $0$ & $0.327^{+0.006}$ &  $25.32$ & $0.0215$ \\
$\mathcal{S}^{2}\times\mathcal{I}$ & $102400$ & $40$ & $3\cdot10^{3}$ & $0.1613$ & $0.3902_{-0.0001}^{+0.0001}$ & $141.32$ & $0.0459$ \\
\bottomrule
\end{tabularx}
\caption{The goodness of fit, as measured by the chi-squared per degree of freedom $\chi_{\mathrm{pdf}}^{2}$, of the function \eqref{discretedSvolprofile} to $\langle N_{2}^{SL}(\tau)\rangle$ for a variety of ensembles of $N(\mathcal{T}_{c})$ causal triangulations characterized by the spacetime topology, the target number $\bar{N}_{3}$ of $3$-simplices, the number $T$ of time slices, the fractional temporal extent $\Delta\tau_{\mathrm{stalk}}/T$ of the stalk, the best fit value of the fit parameter $\tilde{s}_{0}$, and the error $\epsilon(\tilde{s}_{0})$ in the fit parameter $\tilde{s}_{0}$. If no error $\epsilon(\tilde{s}_{0})$ is reported, then $\epsilon(\tilde{s}_{0})<10^{-6}$.}
\label{goodnessoffit}
\end{table}
We draw two conclusions from the values of $\chi^{2}_{\mathrm{pdf}}$. First, at fixed target number $\bar{N}_{3}$ of $3$-simplices and coupling $k_{0}$, there is a particular number $T^{*}$ of time slices that optimizes the fit of the function \eqref{discretedSvolprofile} to $\langle N_{2}^{SL}(\tau)\rangle$. For instance, we find that $T^{*}\approx 21$ for $\bar{N}_{3}=30850$ and $k_{0}=1.00$. This optimization occurs for an ensemble of causal triangulations characterized by a sufficiently small $T$ such that stalks are absent. As we increase $T$ above $T^{*}$, gradually developing stalks, the goodness of fit first worsens for sufficiently short stalks and then improves for sufficiently long stalks. Second, one can achieve an equivalent goodness of fit for a notably smaller target number $\bar{N}_{3}$ of $3$-simplices by working with the optimal number $T^{*}$ of time slices. The values of the chi-squared per degree of freedom $\chi^{2}_{\mathrm{pdf}}$ normalized by the number $N(\mathcal{T}_{c})$ of causal triangulations make this particularly evident. This fact could prove useful for future statistically detailed studies of causal dynamical triangulations.




\subsection{Minimal initial boundary and nonminimal final boundary}

\begin{figure}[!ht]
\centering
\includegraphics[scale=1.3]{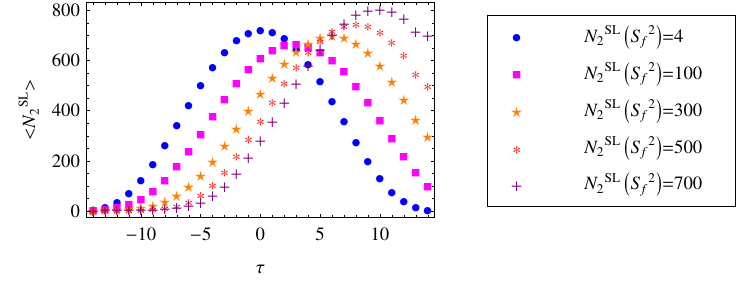}
\caption[Optional caption for list of figures]{Ensemble average number $\langle N_{2}^{SL}\rangle$ of spacelike $2$-simplices as a function of the discrete time coordinate $\tau$ for $\bar{N}_{3}=30850$ and $k_{0}=1.0$.}
\label{HHwavefunction}
\end{figure}

We next take the initial boundary $\mathcal{S}_{i}^{2}$ as the minimal triangulation of the $2$-sphere and the final boundary $\mathcal{S}_{f}^{2}$ as a nonminimal triangulation of the $2$-sphere. In particular, for $T=29$, $\bar{N}_{3}=30850$, and $k_{0}=1.00$, we consider transition amplitudes to final boundaries with $N_{2}^{SL}(\mathcal{S}^{2}_{f})\in\{4,100,300,500,700\}$. (The first such transition amplitude is of course one of those considered previously.) In figure \ref{HHwavefunction} we display $\langle N_{2}^{SL}(\tau)\rangle$ for each of these transition amplitudes. 

We wish to determine if the plots of $\langle N_{2}^{SL}(\tau)\rangle$ in figure \ref{HHwavefunction} accord with the semiclassical expectations discussed above, which suggest that each transition amplitude represented in figure \ref{HHwavefunction} should be dominated by a portion of Euclidean de Sitter spacetime. (Clearly, these transition amplitudes have $\rho_{f}<l_{dS}$.) If this is the case, then we expect the function \eqref{discretedSvolprofile} to fit $\langle N_{2}^{SL}(\tau)\rangle$ for a restriction on the upper limit of the discrete time coordinate $\tau$. While the function \eqref{discretedSvolprofile} fits exceedingly well to $\langle N_{2}^{SL}(\tau)\rangle$ for $N_{2}^{SL}(\mathcal{S}_{f}^{2})=4$, we readily observe that this does not naively hold true for $N_{2}^{SL}(\mathcal{S}_{f}^{2})>4$. If, however, we identify the early time behavior of $\langle N_{2}^{SL}(\tau)\rangle$ as that of a stalk, then we can test whether or not a restriction of the function \eqref{discretedSvolprofile} fits the late time behavior of $\langle N_{2}^{SL}(\tau)\rangle$. 

\begin{figure}[!ht]
\centering
\subfigure[ ]{
\includegraphics[scale=0.405]{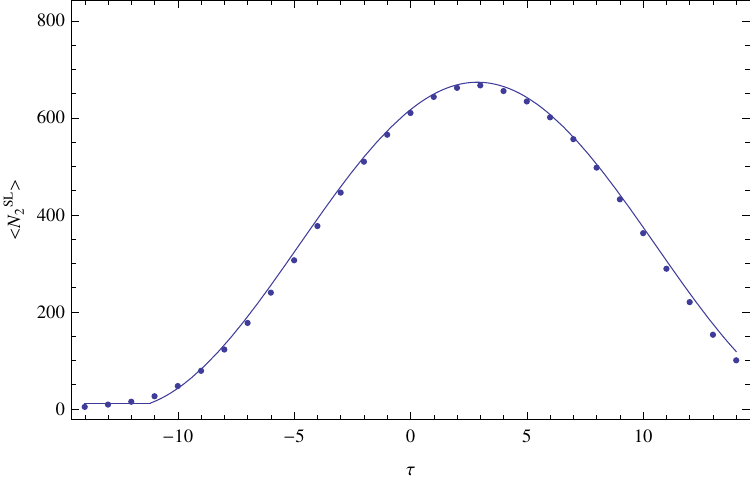}
\label{dSvolfit28slicesFB100}
}
\subfigure[ ]{
\includegraphics[scale=0.405]{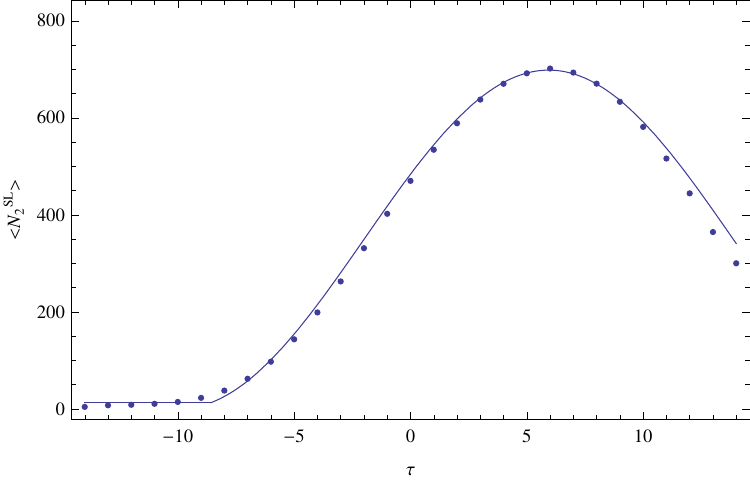}
\label{dSvolfit28slicesFB300}
}\\
\subfigure[ ]{
\includegraphics[scale=0.405]{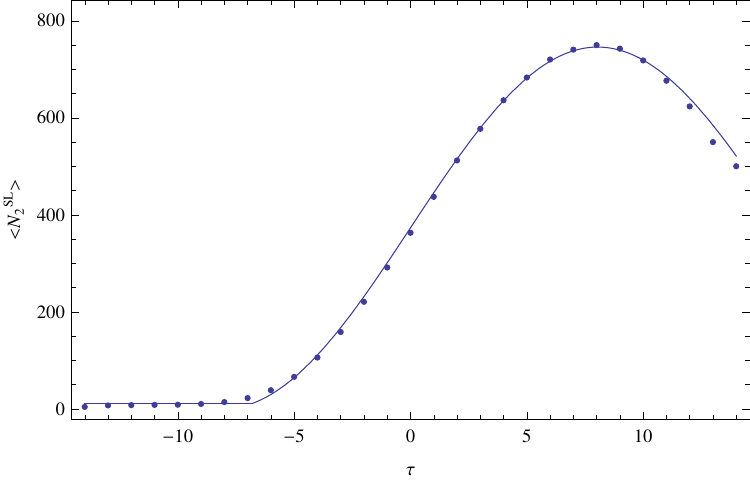}
\label{dSvolfit28slicesFB500}
}
\subfigure[ ]{
\includegraphics[scale=0.405]{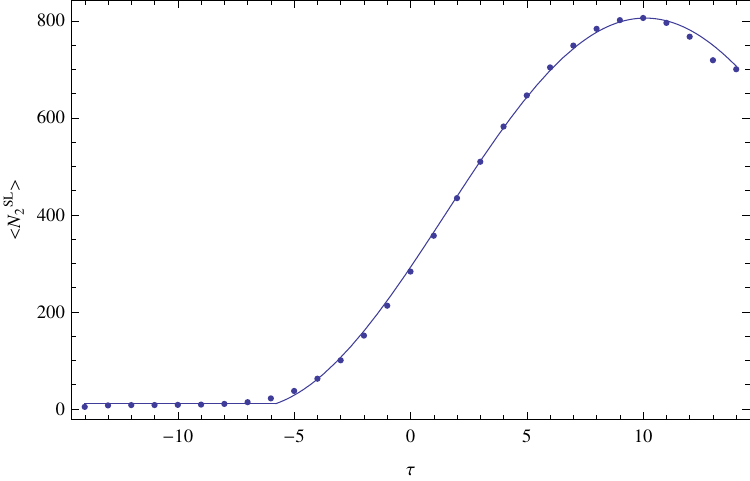}
\label{dSvolfit28slicesFB700}
}
\caption[Optional caption for list of figures]{Ensemble average number $\langle N_{2}^{SL}\rangle$ of spacelike $2$-simplices as a function of the discrete time coordinate $\tau$ for $T=29$, $\bar{N}_{3}=30850$ and $k_{0}=1.00$. \subref{dSvolfit28slicesFB100} $N_{2}^{SL}(\mathcal{S}_{f}^{2})=100$ \subref{dSvolfit28slicesFB300} $N_{2}^{SL}(\mathcal{S}_{f}^{2})=300$  \subref{dSvolfit28slicesFB500} $N_{2}^{SL}(\mathcal{S}_{f}^{2})=500$  \subref{dSvolfit28slicesFB700} $N_{2}^{SL}(\mathcal{S}_{f}^{2})=700$}
\label{minnonmin}
\end{figure}

There still remains one further issue to address: in attempting these fits, what value do we take for the ensemble average number $\langle N_{3}^{(1,3)}\rangle$ of $(1,3)$ $3$-simplices, which sets the overall scaling of the function \eqref{discretedSvolprofile}? The value of $\langle N_{3}^{(1,3)}\rangle$ appearing in the function \eqref{discretedSvolprofile} represents the total number of $(1,3)$ 3-simplices within all of Euclidean de Sitter spacetime. We thus propose taking $\langle N_{3}^{(1,3)}\rangle$ as the effective ensemble average number $\langle N_{3}^{(1,3)}\rangle_{\mathrm{eff}}$ of $(1,3)$ $3$-simplices that would be present without the restriction on the upper limit of $\tau$. We compute $\langle N_{3}^{(1,3)}\rangle_{\mathrm{eff}}$ as twice the number of $(1,3)$ $3$-simplices to the past of the time slice with maximal discrete spatial $2$-volume. In figure \ref{minnonmin} we display the fits of the function \eqref{discretedSvolprofile} to $\langle N_{2}^{SL}(\tau)\rangle$ for each of these transition amplitudes.

\begin{figure}[!ht]
\centering
\subfigure[ ]{
\includegraphics[scale=0.405]{dSvolfit28slicesFB500.pdf}
\label{dSvolfit28slicesFB500}
}
\subfigure[ ]{
\includegraphics[scale=0.405]{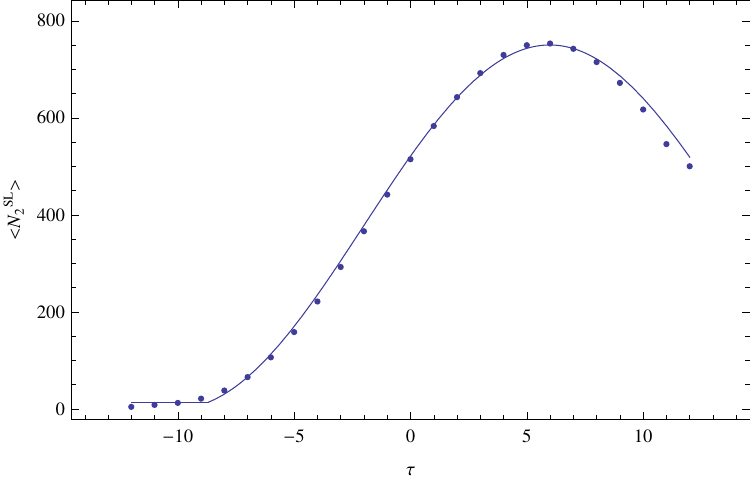}
\label{dSvolfit24slicesFB500}
}
\subfigure[ ]{
\includegraphics[scale=0.405]{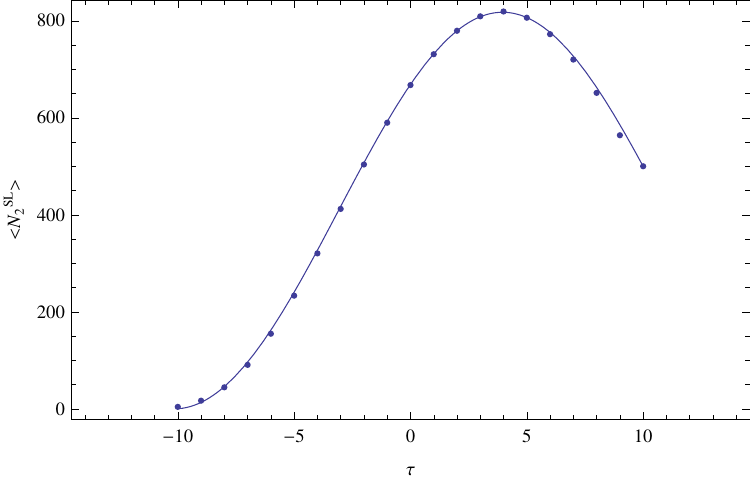}
\label{dSvolfit20slicesFB500}
}
\caption[Optional caption for list of figures]{Ensemble average number $\langle N_{2}^{SL}\rangle$ of spacelike $2$-simplices as a function of the discrete time coordinate $\tau$ for $\bar{N}_{3}=30850$, $k_{0}=1.00$, and $N_{2}^{SL}(\mathcal{S}^{2}_{f})=500$. \subref{dSvolfit28slicesFB500} $T=29$ \subref{dSvolfit24slicesFB500} $T=25$ \subref{dSvolfit20slicesFB500} $T=21$}
\label{varyingTminnonmin}
\end{figure}

Although the function \eqref{discretedSvolprofile} fits $\langle N_{2}^{SL}(\tau)\rangle$ rather well, there is a disconcerting deviation in the fit close to the final boundary. The deviation grows with the discrete spatial $2$-volume $N_{2}^{SL}(\mathcal{S}_{f}^{2})$ of the final boundary, which also corresponds to a growth in the temporal extent of the stalk. To explore the origin of this deviation, we consider the transition amplitudes for $N_{2}^{SL}(\mathcal{S}^{2}_{f})=500$ for two successively smaller numbers $T$ of time slices keeping $\bar{N}_{3}=30850$ and $k_{0}=1.00$. In figure \ref{varyingTminnonmin} we display $\langle N_{2}^{SL}(\tau)\rangle$ for $T=29$ (as above), $T=25$, and $T=21$ fit to the function \eqref{discretedSvolprofile}.

Clearly, as the temporal extent of the stalk diminishes, the deviation in the fit close to the final boundary also diminishes. Based on these findings, we suggest the following interpretation. The deviation in the fit close to the final boundary indicates the beginning of the formation of another stalk, albeit composed of time slices with nonminimal spatial extent. This possibility is consistent with the behavior exhibited in figure \ref{minnonmin}, particularly for the case of $N_{2}^{SL}(\mathcal{S}_{f}^{2})=700$ and in figure \ref{varyingTminnonmin}. Moreover, in the cases of periodic temporal boundary conditions and of minimal initial and final boundary conditions, when a stalk has formed, there are transitional regions at the junctions with the central accumulation of discrete spatial $2$-volume where the fit of the function \eqref{discretedSvolprofile} deviates in a similar manner. See, for instance, figures \ref{comparisons}\subref{periodicdSvolfit64slices} and \ref{minminvaryingTstalks}. We are currently investigating further this phenomenon and its proffered explanation.

As the plots of $\langle N_{2}^{SL}(\tau)\rangle$ in figures \ref{minnonmin} and \ref{varyingTminnonmin} demonstrate, the partition function \eqref{partitionfunctionfixedTN} for minimal initial boundary and nonminimal final boundary is dominated by the extremum of the action \eqref{completeEaction} corresponding to the portion of Euclidean de Sitter spacetime containing more than half of the discrete spacetime $3$-volume of Euclidean de Sitter spacetime. Interestingly, Hartle and Hawking's no-boundary proposal predicts just the opposite: that the portion of Euclidean de Sitter spacetime containing less than half of the spacetime $3$-volume should dominate the path integral \cite{JBH&SWH}. Halliwell and Louko pointed out that the proposal of Hartle and Hawking is not unique: there are other consistent choices of steepest descent integration contours, one of which yields the result that we have obtained \cite{JJH&JL}. We are currently attempting to determine to which analytic minisuperspace quantization the technique of causal dynamical triangulations corresponds. We hope that this investigation illuminates the relation of the causal dynamical triangulations quantization scheme to other proposals for the quantization of classical theories of gravity.


\subsection{Nonminimal initial and final boundaries}

\begin{figure}[!ht]
\centering
\subfigure[ ]{
\includegraphics[scale=0.405]{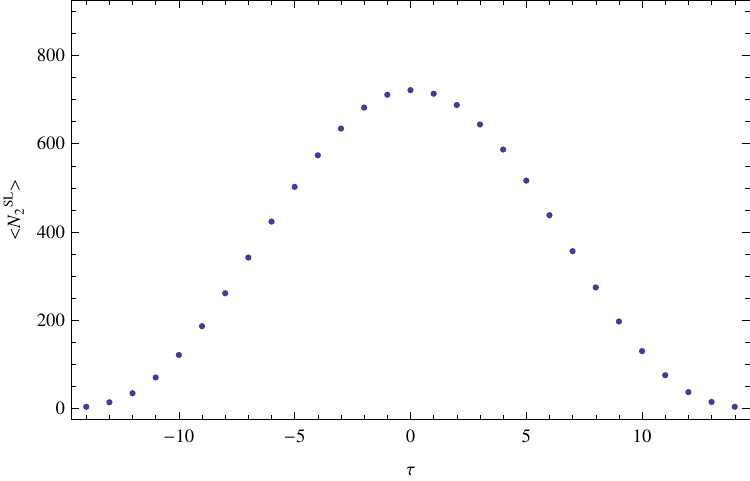}
\label{volprofIB4FB4}
}
\subfigure[ ]{
\includegraphics[scale=0.405]{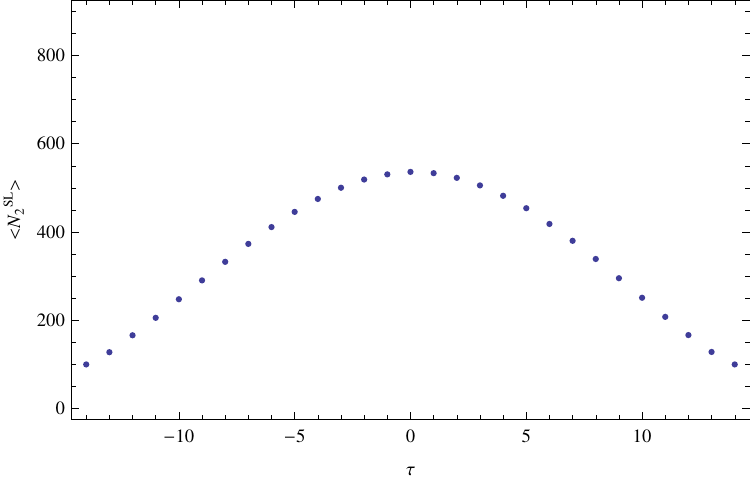}
\label{volprofIB100FB100}
}\\
\subfigure[ ]{
\includegraphics[scale=0.405]{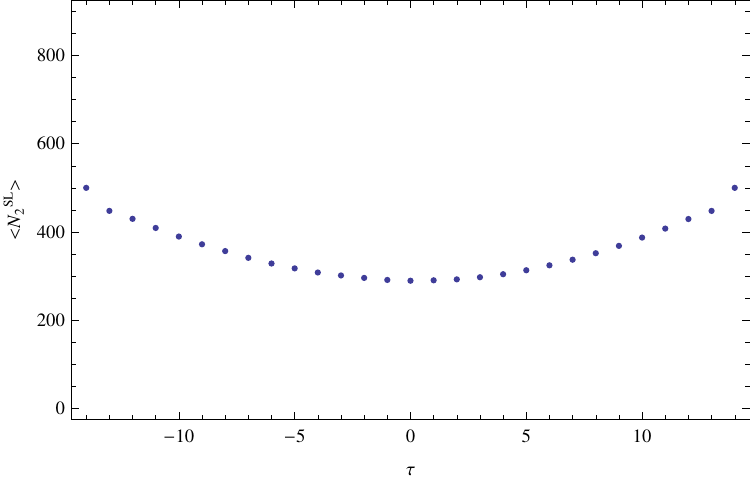}
\label{volprofIB500FB500}
}
\subfigure[ ]{
\includegraphics[scale=0.405]{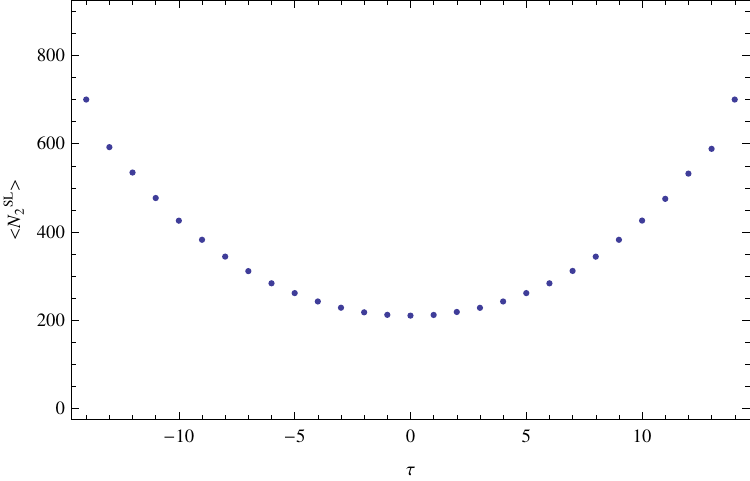}
\label{volprofIB700FB700}
}
\subfigure[ ]{
\includegraphics[scale=0.405]{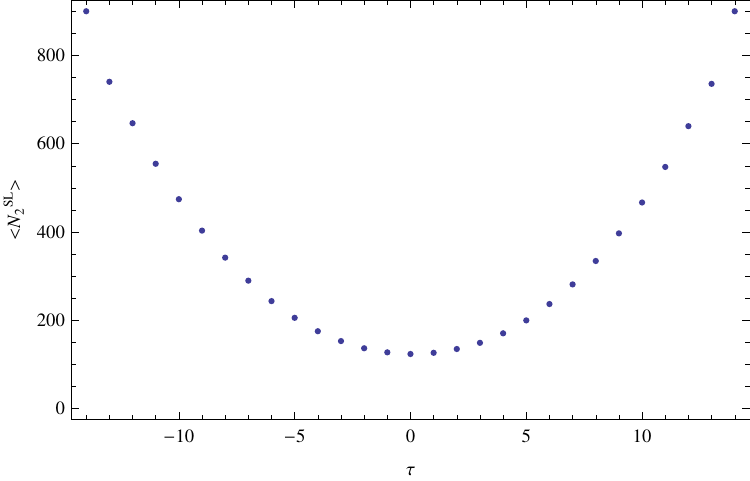}
\label{volprofIB900FB900}
}
\caption[Optional caption for list of figures]{Ensemble average number $\langle N_{2}^{SL}\rangle$ of spacelike $2$-simplices as a function of the discrete time coordinate $\tau$ for $T=29$, $\bar{N}_{3}=30850$, and $k_{0}=1.00$. \subref{volprofIB4FB4} $N_{2}^{SL}(\mathcal{S}_{i}^{2})=4$, $N_{2}^{SL}(\mathcal{S}_{f}^{2})=4$  \subref{volprofIB100FB100} $N_{2}^{SL}(\mathcal{S}_{i}^{2})=100$, $N_{2}^{SL}(\mathcal{S}_{f}^{2})=100$ \subref{volprofIB500FB500} $N_{2}^{SL}(\mathcal{S}_{i}^{2})=500$, $N_{2}^{SL}(\mathcal{S}_{f}^{2})=500$  \subref{volprofIB700FB700} $N_{2}^{SL}(\mathcal{S}_{i}^{2})=700$, $N_{2}^{SL}(\mathcal{S}_{f}^{2})=700$ \subref{volprofIB900FB900} $N_{2}^{SL}(\mathcal{S}_{i}^{2})=900$, $N_{2}^{SL}(\mathcal{S}_{f}^{2})=900$}
\label{nonminnonminsame}
\end{figure}

We finally take both the initial boundary $\mathcal{S}_{i}^{2}$ and the final boundary $\mathcal{S}_{f}^{2}$ as nonminimal triangulations of the $2$-sphere. In particular, for $T=29$, $\bar{N}_{3}=30850$, and $k_{0}=1.00$, we first consider transition amplitudes characterized by $N_{2}^{SL}(\mathcal{S}_{i}^{2})=N_{2}^{SL}(\mathcal{S}_{f}^{2})=4$, $N_{2}^{SL}(\mathcal{S}_{i}^{2})=N_{2}^{SL}(\mathcal{S}_{f}^{2})=100$, $N_{2}^{SL}(\mathcal{S}_{i}^{2})=N_{2}^{SL}(\mathcal{S}_{f}^{2})=500$, $N_{2}^{SL}(\mathcal{S}_{i}^{2})=N_{2}^{SL}(\mathcal{S}_{f}^{2})=700$, and $N_{2}^{SL}(\mathcal{S}_{i}^{2})=N_{2}^{SL}(\mathcal{S}_{f}^{2})=900$. (The first such transition amplitude is of course one of those considered previously.) In figure \ref{nonminnonminsame} we display $\langle N_{2}^{SL}(\tau)\rangle$ for each of these five transition amplitudes. 

\begin{figure}[!ht]
\centering
\subfigure[ ]{
\includegraphics[scale=0.405]{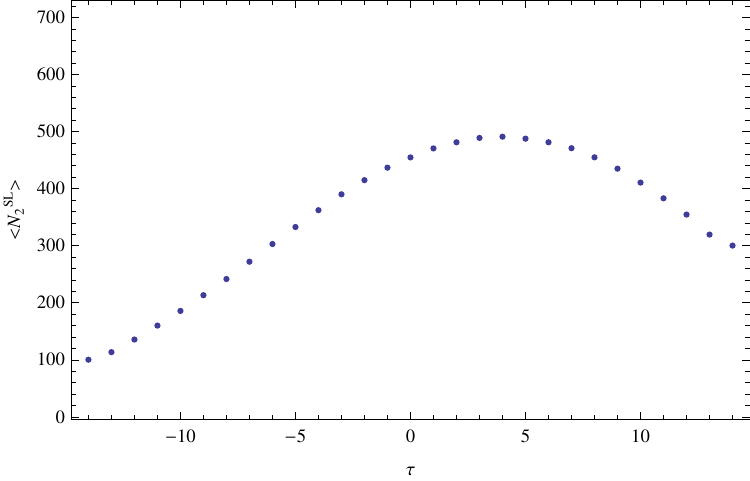}
\label{volprofileIB100FB300}
}
\subfigure[ ]{
\includegraphics[scale=0.405]{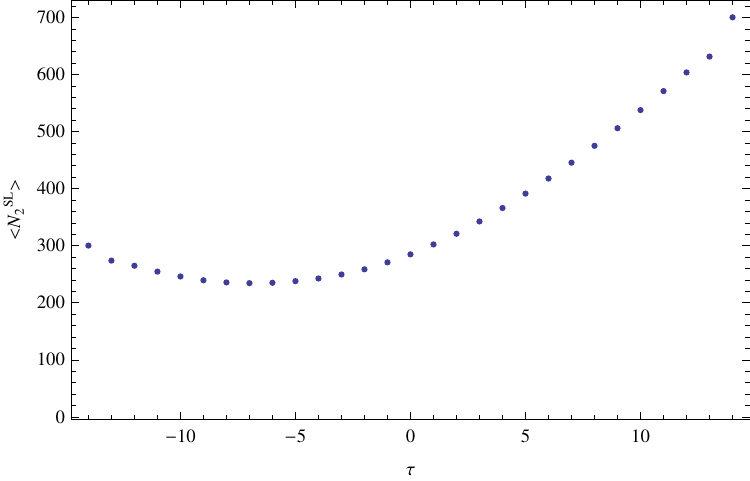}
\label{volprofileIB300FB700}
}
\subfigure[ ]{
\includegraphics[scale=0.405]{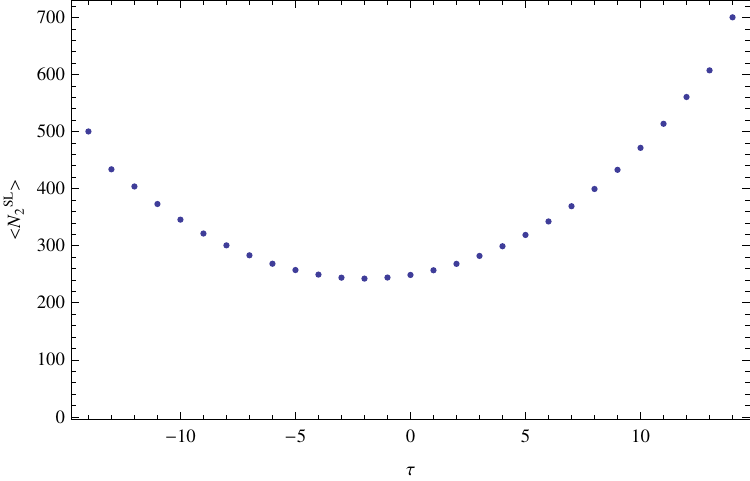}
\label{volprofileIB500FB700}
}
\caption[Optional caption for list of figures]{Ensemble average number $\langle N_{2}^{SL}\rangle$ of spacelike $2$-simplices as a function of the discrete time coordinate $\tau$ for $T=29$, $\bar{N}_{3}=30850$, and $k_{0}=1.00$. \subref{volprofileIB100FB300} $N_{2}^{SL}(\mathcal{S}_{i}^{2})=100$, $N_{2}^{SL}(\mathcal{S}_{f}^{2})=300$  \subref{volprofileIB300FB700} $N_{2}^{SL}(\mathcal{S}_{i}^{2})=300$, $N_{2}^{SL}(\mathcal{S}_{f}^{2})=700$ \subref{volprofileIB500FB700} $N_{2}^{SL}(\mathcal{S}_{i}^{2})=500$, $N_{2}^{SL}(\mathcal{S}_{f}^{2})=700$}
\label{nonminnonmin}
\end{figure}

Also for $T=29$, $\bar{N}_{3}=30850$, and $k_{0}=1.00$, we consider the transition amplitudes characterized by $\{N_{2}^{SL}(\mathcal{S}_{i}^{2})=100,N_{2}^{SL}(\mathcal{S}_{f}^{2})=300\}$, $\{N_{2}^{SL}(\mathcal{S}_{i}^{2})=300,N_{2}^{SL}(\mathcal{S}_{f}^{2})=700\}$, and $\{N_{2}^{SL}(\mathcal{S}_{i}^{2})=500,N_{2}^{SL}(\mathcal{S}_{i}^{2})=700\}$. In figure \ref{nonminnonmin} we display $\langle N_{2}^{SL}(\tau)\rangle$ for each of these three transition amplitudes. 

\begin{figure}[!ht]
\centering
\subfigure[ ]{
\includegraphics[scale=0.405]{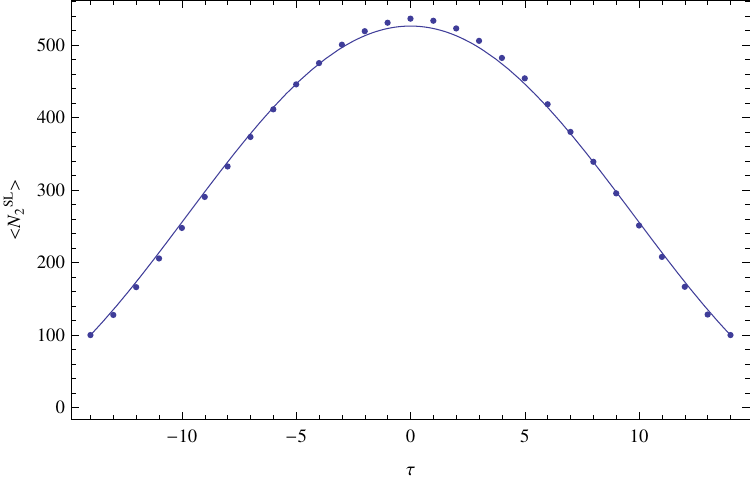}
\label{volproffitIB100FB100}
}
\subfigure[ ]{
\includegraphics[scale=0.405]{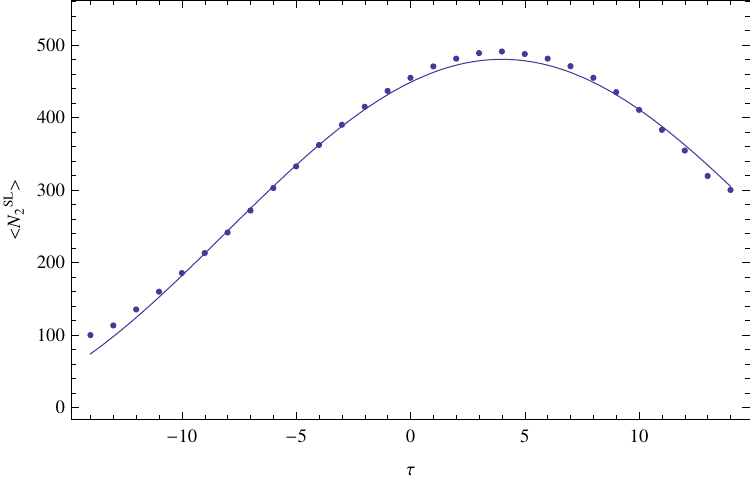}
\label{volproffitIB100FB300}
}
\caption[Optional caption for list of figures]{Ensemble average number $\langle N_{2}^{SL}\rangle$ of spacelike $2$-simplices as a function of the discrete time coordinate $\tau$ for $T=29$, $\bar{N}_{3}=30850$, and $k_{0}=1.00$. \subref{volproffitIB100FB100} $N_{2}^{SL}(\mathcal{S}_{i}^{2})=100$, $N_{2}^{SL}(\mathcal{S}_{f}^{2})=100$ \subref{volproffitIB100FB300} $N_{2}^{SL}(\mathcal{S}_{i}^{2})=100$, $N_{2}^{SL}(\mathcal{S}_{f}^{2})=300$}
\label{nonminnonminvolproffits}
\end{figure}

Clearly, for the transition amplitudes represented in figures \ref{nonminnonminsame} and \ref{nonminnonmin}, $\langle N_{2}^{SL}(\tau)\rangle$ only appears to conform to a portion of Euclidean de Sitter spacetime if the sum $N_{2}^{SL}(\mathcal{S}_{i}^{2})+N_{2}^{SL}(\mathcal{S}_{f}^{2})$ is sufficiently small. We confirm this quantitatively by fitting the function \eqref{discretedSvolprofile} to $\langle N_{2}^{SL}(\tau)\rangle$ for two such cases. In performing these fits, we again require the appropriate values for the ensemble average number $\langle N_{3}^{(1,3)}\rangle$ of $(1,3)$ $3$-simplices. Instead of proceeding as in the previous subsection, we circumvent this issue by employing the variant of the function \eqref{discretedSvolprofile} appropriate to a portion of Euclidean de Sitter spacetime. We derive this variant in appendix \ref{volumeprofilederivation}; we could have elected to use this variant in the previous subsection as well. We display the results of these fits in figure \ref{nonminnonminvolproffits}. 
In the case of $\{N_{2}^{SL}(\mathcal{S}_{i}^{2})=100,N_{2}^{SL}(\mathcal{S}_{f}^{2})=300\}$, there is clearly a deviation in the fit close to both the initial and final boundaries similar to what we found in the previous subsection. 

If the sum $N_{2}^{SL}(\mathcal{S}_{i}^{2})+N_{2}^{SL}(\mathcal{S}_{f}^{2})$ is insufficiently small, then $\langle N_{2}^{SL}(\tau)\rangle$ exhibits a markedly different dynamics, possibly hyperbolic in nature. 
For the chosen values of $T$, $\bar{N}_{3}$, and $k_{0}$, there is a transition from the typical dynamics of phase C to the possibly hyperbolic dynamics when $N_{2}^{SL}(\mathcal{S}_{i}^{2})+N_{2}^{SL}(\mathcal{S}_{f}^{2})\approx 600$. We have determined the approximate location of this transition by additionally considering the transition amplitudes characterized by $N_{2}^{SL}(\mathcal{S}_{i}^{2})=N_{2}^{SL}(\mathcal{S}_{f}^{2})=300$ and by $\{N_{2}^{SL}(\mathcal{S}_{i}^{2})=100,N_{2}^{SL}(\mathcal{S}_{f}^{2})=500\}$, both of which show a mixture of these two dynamical behaviors. 

Our above discussion of semiclassical expectations for these transition amplitudes suggests an interpretation of these results. The ensembles characterized by $N_{2}^{SL}(\mathcal{S}_{i}^{2})=N_{2}^{SL}(\mathcal{S}_{f}^{2})=100$ and by $\{N_{2}^{SL}(\mathcal{S}_{i}^{2})=100,N_{2}^{SL}(\mathcal{S}_{f}^{2})=300\}$, which exhibit the typical dynamics of phase C, correspond to the regime in which both the initial and final scale factors are less than the de Sitter length for the given discrete spacetime $3$-volume. This is the situation depicted in figure \ref{extrema}\subref{portionofsphere}. Presumably then, the other five ensembles considered here, which exhibit possibly hyperbolic dynamics, correspond to the regime in which both the initial and final scale factors exceed the de Sitter length for the given discrete spacetime $3$-volume. Recall that in this case the amplitude for the scale factor oscillates within an envelope consistent with the probability distribution of scale factors in Lorentzian de Sitter spacetime. Naively, this appears to be the result that we have achieved: $\langle N_{2}^{SL}(\tau)\rangle$ appears to conform to a portion of Lorentzian de Sitter spacetime. We are currently investigating whether or not this is quantitatively the case. 

Supposing that a discretization of a portion of the spatial $2$-volume as a function of the global time coordinate of Lorentzian de Sitter spacetime does indeed fit well to $\langle N_{2}^{SL}(\tau)\rangle$, what conclusions should we draw? This finding would imply that the quantization scheme of causal dynamical triangulations functions similarly to that of the no-boundary proposal of Hartle and Hawking: one obtains transition amplitudes dominated by Lorentzian spacetimes \emph{via} a Euclidean path integral \cite{JBH&SWH}. Possibly then, as in the no-boundary proposal, Lorentzian not Euclidean de Sitter spacetime dominates the ground state of quantum spacetime geometry on sufficiently large scales. Regardless of the viability of these conclusions, the latter five transition amplitudes appear to represent numerical simulations of portions of temporally unbounded quantum spacetime geometry, a first within the causal dynamical triangulations approach.

\section{Conclusion}\label{conclusion}

We have initiated an investigation of transition amplitudes within the causal dynamical triangulations of $(2+1)$-dimensional Einstein gravity with positive cosmological constant. We have focused on transition amplitudes from a past spacelike hypersurface of fixed intrinsic geometry to a future spacelike hypersurface of fixed intrinsic geometry, these hypersurfaces being leaves of the foliation of causal triangulations. Our study thus required a generalization of previous numerical investigations of causal dynamical triangulations to the setting of fixed initial and final boundary geometries. 

Specifically, we explored three classes of such transition amplitudes for spatial topology of a $2$-sphere: minimal boundary--minimal boundary, minimal boundary--nonminimal boundary, and nonminimal boundary--nonminimal boundary. When averaged over all geometrical degrees of freedom of each boundary except for the discrete spatial $2$-volume, these transition amplitudes are all evidently compatible with the gravitational effective action \eqref{effectiveaction} previously demonstrated to describe the ground state of phase C. We are currently investigating the extent to which this compatibility holds quantitatively at the level of deviations from the ensemble average.

Each class of transition amplitude has afforded new insights into the approach of causal dynamical triangulations. The minimial boundary--minimal boundary transition amplitudes, as well as the minimal boundary--nonminimal boundary transition amplitudes, definitively demonstrate that the stalk regions of quantum spacetime geometry, previously observed in all numerical simulations within phase C, are indeed artifacts of the numerical implementation. Nevertheless, we are interested in understanding the formation of stalks, particularly as they appear in the minimal boundary--nonminimal boundary transition amplitudes. To this end we are exploring whether the minisuperspace model defined by the effective action \eqref{effectiveaction} can explain the dynamical formation of stalks. 

The minimal boundary--nonminimal boundary transition amplitudes---and possibly the nonminimal\linebreak boundary--nonminimal boundary transition amplitudes---provide for direct comparisons of the quantizations of causal dynamical triangulations and of certain continuum approaches. In particular, these transition amplitudes indicate that the technique of causal dynamical triangulations does not correspond precisely to that of Hartle and Hawking's no-boundary proposal. We are currently attempting to ascertain whether the quantization scheme of causal dynamical triangulations corresponds to one of the variants of this proposal put forward by Halliwell and Louko \cite{JJH&JL}. We also wish to determine whether the nonminimal boundary--nonminimal boundary transition amplitudes---apparently representing numerical simulations of portions of temporally unbounded quantum spacetime geometry---match the quantitative expectations of these scenarios. 

Our study of transition amplitudes opens the door to a multitude of interesting new explorations of causal dynamical triangulations. As discussed above, we have only studied geometry-averaged transition amplitudes, that is, those depending solely on the discrete spatial $2$-volumes of the bounding geometries. Presumably, there is considerably more information within transition amplitudes that probe the full dependence on the boundaries' intrinsic geometries. To study such transition amplitudes, one would require the ability first to characterize the geometrical degrees of freedom of triangulated boundary geometries and then to design triangulated boundary geometries with chosen characteristics. Some of the techniques employed by Sachs, in combination with our Markov chain Monte Carlo algorithm for generating randomly triangulated $2$-spheres, may serve this purpose \cite{MKS}. We are particularly interested in the possibility of observing effects that are not describable within a minisuperspace truncation of the metric degrees of freedom. Specifically, by designing appropriate boundary $2$-spheres, we hope to investigate the absence or presence of propagating degrees of freedom in the quantum theory. 


Our techniques might also prove useful in further studying the causal dynamical triangulations of projectable Ho\v{r}ava-Lifshitz gravity as initiated by Anderson \emph{et al} \cite{CA&SJC&JHC&PH&RKK&PZ}. These authors discovered an auxiliary phase of quantum spacetime geometry---their so-called phase E---but they did not have a method for coherently ensemble averaging the discrete spatial $2$-volume as a function of the discrete time coordinate in this phase. By fixing the intrinsic geometries of the initial and final boundaries, one could employ the boundaries to temporally align each causal triangulation in an ensemble, allowing for coherent averaging in phase E. Furthermore, following on the final aspiration of the previous paragraph, one could attempt to determine whether or not there is a propagating scalar mode within this quantization of Ho\v{r}ava-Lifshitz gravity. Such investigations would work towards ascertaining the relationship between the causal dynamical triangulations of Einstein gravity and quantum Ho\v{r}ava-Lifshitz gravity \cite{JA&AG&SJ&JJ&RL}, which have recently been shown to be equivalent in $1+1$ dimensions \cite{JA&LG&YS&YW}.

\section*{Acknowledgments}

We wish to thank Steve Carlip for his guidance and input throughout the course of this project. We are grateful to Christian Anderson, David Kamensky, and especially Rajesh Kommu for allowing us to employ parts of their computer codes. JHC also wishes to thank Kyle Lee for several useful conversations. JHC acknowledges support from the Department of Energy under grant DE-FG02-91ER40674. JMM acknowledges support from the National Science Foundation under REU grant PHY-1004848 at the University of California, Davis. This work utilized the Janus supercomputer, which is supported by the National Science Foundation (award number CNS-0821794) and the University of Colorado, Boulder. The Janus supercomputer is a joint effort of the University of Colorado, Boulder, the University of Colorado, Denver, and the National Center for Atmospheric Research.

\appendix

\section{On the consistency of the action $S_{R}[\mathcal{T}_{c}]$ under composition}\label{composition}

We demonstrate that our prescription \eqref{GHYCDTaction} for the Gibbons-Hawking-York boundary term is consistent with the prescription \eqref{periodicCDTaction} for the Einstein-Hilbert action within the causal dynamical triangulations of $(2+1)$-dimensional Einstein gravity with positive cosmological constant for $2$-sphere spatial topology. We make such a demonstration by verifying that the prescription \eqref{GHYCDTaction} reproduces the prescription \eqref{periodicCDTaction} under the composition of two spacetime regions sharing a common boundary. Hartle and Sorkin employed this criterion in deriving the form \eqref{HSaction} of the Gibbons-Hawking-York boundary term in Regge calculus \cite{JBH&RS}, so we certainly expect our prescription \eqref{GHYCDTaction} to preserve this criterion. 

Consider two causal triangulations $\mathcal{T}_{c}$ and $\mathcal{T}_{c}'$ both with $2$-sphere spatial topology and line interval temporal topology. The boundary $\partial\mathcal{T}_{c}$ consists of an initial $2$-sphere $\mathcal{S}_{i}^{2}$ and a final $2$-sphere $\mathcal{S}_{f}^{2}$, and the boundary $\partial\mathcal{T}_{c}'$ consists of an initial $2$-sphere $\mathcal{S'}_{i}^{2}$ and a final $2$-sphere $\mathcal{S'}_{f}^{2}$. To compose the two causal triangulations $\mathcal{T}_{c}$ and $\mathcal{T}_{c}'$, we take the $2$-spheres $\mathcal{S}_{f}^{2}$ and $\mathcal{S'}_{i}^{2}$ to have the same intrinsic geometry, and we orient these $2$-spheres $\mathcal{S}_{f}^{2}$ and $\mathcal{S'}_{i}^{2}$ to have coincident normal vectors. We may thus identify these two $2$-spheres as $\mathcal{S}_{\mathcal{C}}^{2}$, the common boundary. The Gibbons-Hawking-York boundary term for $\mathcal{S}_{\mathcal{C}}^{2}$ from the two causal triangulations $\mathcal{T}_{c}$ and $\mathcal{T}_{c}'$ is
\begin{eqnarray}
&&\frac{a}{8\pi G}\left[\frac{\pi}{i} N_{1}^{SL}(\mathcal{S}_{f}^{2})-\frac{2}{i}\theta_{SL}^{(1,3)}N_{1}^{SL}(\mathcal{S}_{f}^{2})-\frac{1}{i}\theta_{SL}^{(2,2)}N_{3\downarrow}^{(2,2)}(\mathcal{S}_{f}^{2})\right]\nonumber\\ &&\qquad+\frac{a}{8\pi G}\left[\frac{\pi}{i} N_{1}^{SL}(\mathcal{S'}_{i}^{2})-\frac{2}{i}\theta_{SL}^{(3,1)}N_{1}^{SL}(\mathcal{S'}_{i}^{2})-\frac{1}{i}\theta_{SL}^{(2,2)}N_{3\uparrow}^{(2,2)}(\mathcal{S'}_{i}^{2})\right].
\end{eqnarray}
These two contributions to the Gibbons-Hawking-York boundary term combine to give 
\begin{equation}
\frac{a}{8\pi G}\left[\frac{2\pi}{i}N_{1}^{SL}(\mathcal{S}_{\mathcal{C}}^{2})-\frac{4}{i}\theta_{SL}^{(1,3)}N_{1}^{SL}(\mathcal{S}_{\mathcal{C}}^{2})-\frac{2}{i}\theta_{SL}^{(2,2)}N_{3}^{(2,2)}(\mathcal{S}_{\mathcal{C}}^{2})\right],
\end{equation}
precisely the contribution to the Einstein-Hilbert action stemming from the spacelike hinges on the common boundary $\mathcal{S}_{\mathcal{C}}^{2}$.

\section{On the algorithm for inserting spatial $2$-sphere boundaries}\label{algorithm}

We describe the algorithm for incorporating an arbitrary initial or final spacelike boundary $\mathcal{S}_{\mathcal{B}}^{2}$ of $2$-sphere topology into a minimal causal triangulation. Recall that a minimal causal triangulation consists of minimally triangulated 2-spheres, each the surface of a spatial tetrahedron, for every time slice with adjacent time slices connected by the minimal number of timelike edges. This construction results in there being fourteen 3-simplices between adjacent time slices: four $(3,1)$ $3$-simplices connecting spacelike $2$-simplices in the initial slice to
vertices in the final slice, four $(1,3)$ $3$-simplices connecting spacelike $2$-simplices in the final slice to vertices in the initial slice, and six $(2,2)$ $3$-simplices filling in the gaps between these other $3$-simplices to ensure the correct topology.

\begin{figure}[htb]
\begin{center}
\leavevmode
\includegraphics[scale=0.6]{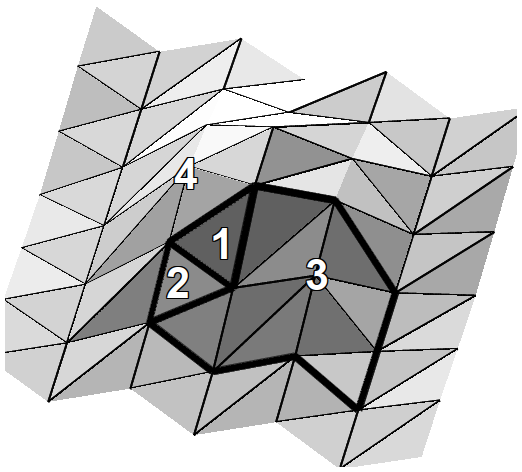}
\caption{An example decomposition of a triangulated spatial $2$-sphere (not completely depicted) into four pseudofaces---labelled $1$, $2$, $3$, and $4$---and the six corresponding pseudoedges.}
\label{fig:pseudo-face-decomposition}
\end{center}
\end{figure}

To insert the arbitrarily triangulated $2$-sphere $\mathcal{S}_{\mathcal{B}}^{2}$ into this triangulation, we first decompose $\mathcal{S}_{\mathcal{B}}^{2}$ into four pseudofaces and six pseudoedges. Each pseudoface is a simply-connected set of spacelike $2$-simplices, and each pseudoedge is a piecewise curve of spacelike $1$-simplices. See figure \ref{fig:pseudo-face-decomposition}. Clearly, each pseudoface is bounded by three pseudoedges. Together the pseudofaces and pseudoedges form a pseudotetrahedron. The decomposition into pseudofaces and pseudoedges is not unique except for the minimally triangulated $2$-sphere. This nonuniqueness does not impact our algorithm, which works for any such decomposition.

We next delete either the initial or final time slice from the minimal causal triangulation. We wish to replace this time slice with the arbitrarily triangulated $2$-sphere $\mathcal{S}_{\mathcal{B}}^{2}$ that we have decomposed into a pseudotetrahedron. To make this replacement, we require a method for correctly connecting $\mathcal{S}_{\mathcal{B}}^{2}$ to the next-to-initial or next-to-final time slice of the minimal causal triangulation. We achieve this connection by replacing the fourteen $3$-simplices formerly connecting the initial and next-to-initial or final and next-to-final time slices with fourteen pseudo-$3$-simplices. In particular, we employ pseudo-$(1,3)$ $3$-simplices, pseudo-$(2,2)$ $3$-simplices, and pseudo-$(3,1)$ $3$-simplices in place of the $(1,3)$ $3$-simplices, $(2,2)$ $3$-simplices, and $(3,1)$ $3$-simplices. A pseudo-$(p,q)$ $3$-simplex is a complex of $(p,q)$ $3$-simplices having the topology of a single $(p,q)$ $3$-simplex and constructed to have the spacelike pseudoedges matching those of the pseudofaces to which it will connect. This algorithm results in a minimal causal triangulation having either an initial or final arbitrarily triangulated $2$-sphere boundary $\mathcal{S}_{\mathcal{B}}^{2}$.


\begin{figure}[htb]
\centering
\includegraphics[scale=0.4]{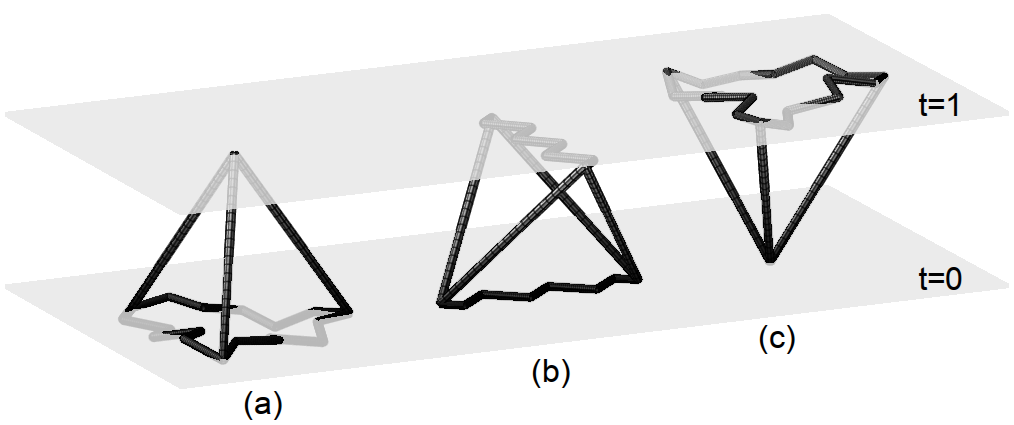}
\caption{The three types of pseudo-$3$-simplices employed in $(2+1)$-dimensional causal dynamical triangulations with fixed $2$-sphere boundaries: (a) pseudo-$(3,1)$ $3$-simplex, (b) pseudo-$(2,2)$ $3$-simplex, (c) pseudo-$(3,1)$ $3$-simplex. The jagged curves are spacelike pseudoedges, and the straight lines are timelike edges.}
\label{fig:pseudo-face-complices}
\end{figure}

\section{A derivation of the function $\mathscr{N}_{2}^{SL}(\tau)$}\label{volumeprofilederivation}

We demonstrate that the function
\begin{equation}\label{discretedSvolprofileapp}
\mathscr{N}_{2}^{SL}(\tau)=\frac{2}{\pi}\frac{\langle N_{3}^{(1,3)}\rangle}{\tilde{s}_{0}\langle N_{3}^{(1,3)}\rangle^{1/3}}\cos^{2}{\left(\frac{\tau}{\tilde{s}_{0}\langle N_{3}^{(1,3)}\rangle^{1/3}}\right)},
\end{equation}
previously given in equation \eqref{discretedSvolprofile}, is the discrete analogue of the spatial $2$-volume $V_{2}$ as a function of the global time coordinate $\eta$ of Euclidean de Sitter spacetime. This demonstration relies crucially on application of the appropriate finite size scaling towards the continuum limit. Dimensional analysis suggests that one approaches the continuum limit by taking the lattice spacing $a$ to zero and letting the number $N_{3}$ of $3$-simplices increase without bound while holding the product $a^{3}N_{3}$ constant. That the gravitational effective action \eqref{effectiveaction} describes the ensemble average spacetime geometry on sufficiently large scales supports the conclusion of this naive analysis \cite{JA&AG&JJ&RL1,JA&AG&JJ&RL2,JA&JJ&RL3,JA&JJ&RL4,JA&JJ&RL5,JA&JJ&RL6,DB&JH}. In particular, we expect the condition
\begin{equation}\label{normalization3}
V_{3}=C_{3}a^{3}N_{3},
\end{equation}
relating the continuum spacetime $3$-volume $V_{3}$ to the lattice spacing $a$ and the number $N_{3}$ of $3$-simplices to hold in the continuum limit. Here, $C_{3}$ is the effective discrete spacetime $3$-volume of a $3$-simplex. Accordingly, we expect discrete quantities having the associated dimensions $a^{p}$ to finite size scale towards the continuum as $N_{3}^{-p/3}$. Specifically, we expect the discrete time coordinate $\tau$ to scale as $\tau N_{3}^{-1/3}$, matching onto the dimensionless continuum time coordinate $\eta V_{3}^{-1/3}$, and we expect the number $N_{2}^{SL}$ of spacelike $2$-simplices to scale as $N_{2}^{SL}N_{3}^{-2/3}$, matching onto the dimensionless continuum spatial $2$-volume $V_{2}V_{3}^{-2/3}$. 

With these considerations we are now prepared to derive the function \eqref{discretedSvolprofileapp}. We start from the definition
\begin{equation}
V_{3}=\int_{-\pi l_{dS}/2\sqrt{g_{\eta\eta}}}^{+\pi l_{dS}/2\sqrt{g_{\eta\eta}}}\mathrm{d}\eta\sqrt{g_{\eta\eta}}V_{2}(\eta),
\end{equation}
which determines the spacetime $3$-volume $V_{3}$ of Euclidean de Sitter spacetime, and the definition
\begin{equation}\label{N3def}
N_{3}=N_{3}^{(1,3)}+N_{3}^{(2,2)}+N_{3}^{(3,1)},
\end{equation}
which determines the number $N_{3}$ of $3$-simplices comprising a causal triangulation $\mathcal{T}_{c}$. For spacetime topology $\mathcal{S}^{2}\times\mathcal{S}^{1}$ one may recast the definition \eqref{N3def} as
\begin{equation}
N_{3}=2(1+\xi)\sum_{\tau=1}^{T}N_{2}^{SL}(\tau),
\end{equation}
where
\begin{equation}
\xi=\frac{N_{3}^{(2,2)}}{N_{3}^{(1,3)}+N_{3}^{(3,1)}}.
\end{equation}
The condition \eqref{normalization3} then becomes
\begin{equation}
\int\mathrm{d}\eta\sqrt{g_{\eta\eta}}V_{2}(\eta)=2C_{3}a^{3}(1+\xi)\sum_{\tau=1}^{T}N_{2}^{SL}(\tau),
\end{equation}
yielding the identification 
\begin{equation}\label{N2id}
\mathrm{d}\eta\sqrt{g_{\eta\eta}}V_{2}(\eta)=2C_{3}a^{3}(1+\xi)N_{2}^{SL}(\tau).
\end{equation}
Given the scale factor \eqref{dSscalefactor} describing Euclidean de Sitter spacetime, the spatial $2$-volume as a function of the global time coordinate is
\begin{equation}\label{continuous2volume}
V_{2}(\eta)=\frac{2V_{3}}{\pi l_{dS}}\cos^{2}{\left(\frac{\sqrt{g_{\eta\eta}}}{l_{dS}}\eta\right)}
\end{equation}
in terms of the spacetime $3$-volume $V_{3}=2\pi^{2}l_{dS}^{3}$. Solving equation \eqref{N2id} for $N_{2}^{SL}(\tau)$ and substituting for $V_{2}(\eta)$ from the expression \eqref{continuous2volume}, we obtain the relation
\begin{equation}\label{discrete2volume2}
N_{2}^{SL}(\tau)=\frac{\mathrm{d}\eta\sqrt{g_{\eta\eta}}}{2a^{3}C_{3}(1+\xi)}\frac{2V_{3}}{\pi l_{dS}}\cos^{2}{\left(\frac{\sqrt{g_{\eta\eta}}}{l_{dS}}\eta\right)}.
\end{equation}
Using the condition \eqref{normalization3} again, we may rewrite equation \eqref{discrete2volume2} as
\begin{equation}\label{discrete2volume3}
N_{2}^{SL}(\tau)=\frac{\mathrm{d}\eta\sqrt{g_{\eta\eta}}}{(1+\xi)}\frac{N_{3}}{\pi l_{dS}}\cos^{2}{\left(\frac{\sqrt{g_{\eta\eta}}}{l_{dS}}\eta\right)}.
\end{equation}
Replacing $\eta$ with $\tau$ according to the above scaling correspondence, equation \eqref{discrete2volume3} becomes
\begin{equation}
N_{2}^{SL}(\tau)=\frac{1}{\pi}\frac{\Delta\tau\sqrt{g_{\eta\eta}}V_{3}^{1/3}N_{3}}{l_{dS}N_{3}^{1/3}(1+\xi)}\cos^{2}{\left(\frac{\sqrt{g_{\eta\eta}}}{l_{dS}}\frac{V_{3}^{1/3}\tau}{N_{3}^{1/3}}\right)}.
\end{equation}
Defining 
\begin{equation}
\frac{1}{s_{0}}=\frac{V_{3}^{1/3}\sqrt{g_{\eta\eta}}}{l_{dS}},
\end{equation}
we finally determine that
\begin{equation}
N_{2}^{SL}(\tau)=\frac{1}{\pi}\frac{N_{3}}{s_{0}(1+\xi)N_{3}^{1/3}}\cos^{2}{\left(\frac{\tau}{s_{0}N_{3}^{1/3}}\right)}
\end{equation}
since $\Delta\tau=1$. In terms of $N_{3}^{(3,1)}$ and the modified parameter $\tilde{s}_{0}=2^{1/3}s_{0}(1+\xi)^{1/3}$,
\begin{equation}\label{discretedSvolprofileappen}
N_{2}^{SL}(\tau)=\frac{2}{\pi}\frac{N_{3}^{(3,1)}}{\tilde{s}_{0}\left(N_{3}^{(3,1)}\right)^{1/3}}\cos^{2}{\left[\frac{\tau}{\tilde{s}_{0}\left(N_{3}^{(3,1)}\right)^{1/3}}\right]}.
\end{equation}
The parameter $\tilde{s}_{0}$ is effectively the dimensionless de Sitter radius. As we only expect the relation \eqref{discretedSvolprofileappen} to hold at the level of the ensemble average, we obtain the function \eqref{discretedSvolprofileapp}.

We readily generalize the result \eqref{discretedSvolprofileappen} to a portion of Euclidean de Sitter spacetime. We now start from the definition
\begin{equation}
V_{3}=\int_{\eta_{i}}^{\eta_{f}}\mathrm{d}\eta\sqrt{g_{\eta\eta}}V_{2}(\eta),
\end{equation}
which determines the spacetime $3$-volume $V_{3}$ of the portion of Euclidean de Sitter spacetime between global times $\eta_{i}$ and $\eta_{f}$ as
\begin{equation}
V_{3}=2\pi l_{dS}^{2}\sqrt{g_{\eta\eta}}\left\{\eta_{f}-\eta_{i}+\frac{l_{dS}}{\sqrt{g_{\eta\eta}}}\sin{\left[\frac{\sqrt{g_{\eta\eta}}(\eta_{f}-\eta_{i})}{l_{dS}}\right]}\cos{\left[\frac{\sqrt{g_{\eta\eta}}(\eta_{f}+\eta_{i})}{l_{dS}}\right]}\right\}.
\end{equation}
Proceeding precisely as above, we eventually find that
\begin{equation}\label{variantdiscretedSvolprofileappen}
N_{2}^{SL}(\tau)=\frac{2N_{3}^{(1,3)}\cos^{2}{\left[\frac{\tau}{\tilde{s}_{0}\left(N_{3}^{(3,1)}\right)^{1/3}}\right]}}{\tau_{f}-\tau_{i}+\tilde{s}_{0}\left(N_{3}^{(1,3)}\right)^{1/3}\sin{\left[\frac{\tau_{f}-\tau_{i}}{\tilde{s}_{0}\left(N_{3}^{(1,3)}\right)^{1/3}}\right]}\cos{\left[\frac{\tau_{f}+\tau_{i}}{\tilde{s}_{0}\left(N_{3}^{(1,3)}\right)^{1/3}}\right]}}
\end{equation}
for the values $\tau_{i}$ and $\tau_{f}$ of the discrete time coordinate corresponding to $\eta_{i}$ and $\eta_{f}$. 

We fit the function \eqref{discretedSvolprofileapp} to the coherent ensemble average number $\langle N_{2}^{SL}(\tau)\rangle$ of spacelike $2$-simplices as a function of the discrete time coordinate, measured directly from Markov chain Monte Carlo data, as follows. In the case of periodic boundary conditions and for minimal boundary--minimal boundary transition amplitudes when stalks are present, we employ in particular the function
\begin{equation}\label{discretedSvolprofileminmin}
\mathscr{N}_{2}^{SL}(\tau)=\left\{\begin{array}{llc}
A & \mathrm{for} & -\frac{T}{2}\leq\tau< -\tau_{m} \\
\frac{2}{\pi}\frac{\langle N_{3}^{(1,3)}\rangle}{\tilde{s}_{0}\langle N_{3}^{(1,3)}\rangle^{1/3}}\cos^{2}{\left(\frac{\tau}{\tilde{s}_{0}\langle N_{3}^{(1,3)}\rangle^{1/3}}\right)} & \mathrm{for} & -\tau_{m}\leq\tau\leq\tau_{m} \\
A & \mathrm{for} & \tau_{m}<\tau\leq\frac{T}{2}
\end{array}\right.
\end{equation}
for the value
\begin{equation}
\tau_{m}=\tilde{s}_{0}\langle N_{3}^{(1,3)}\rangle^{1/3}\cos^{-1}\sqrt{\frac{\pi A\tilde{s}_{0}}{2\langle N_{3}^{(1,3)}\rangle^{2/3}}}
\end{equation}
of the discrete time coordinate that matches the discrete spatial $2$-volume $A$ of the stalk to that of the central accumulation. For minimal boundary--minimal boundary transition amplitudes when stalks are not present, we employ the function \eqref{discretedSvolprofileminmin} with $\tau_{m}=\frac{T}{2}$, and we enforce that the function \eqref{discretedSvolprofileminmin} passes through the boundary values of $\langle N_{2}^{SL}(\tau)\rangle$. The value of $\tilde{s}_{0}$ is thus determined to be the root of the transcendental equation
\begin{equation}
4=\frac{2}{\pi}\frac{\langle N_{3}^{(1,3)}\rangle}{\tilde{s}_{0}\langle N_{3}^{(1,3)}\rangle^{1/3}}\cos^{2}{\left(\frac{\frac{T}{2}}{\tilde{s}_{0}\langle N_{3}^{(1,3)}\rangle^{1/3}}\right)}
\end{equation}
since the boundary value of $\langle N_{2}^{SL}(\tau)\rangle$ is precisely $4$ in this case. We determine whether or not stalks are present by considering both of the above fits, taking the goodness of the fit, described below, as our indicator. 

For minimal boundary--nonminimal boundary transition amplitudes when stalks are present, we employ in particular the function 
\begin{equation}\label{discretedSvolprofileminnonmin}
\mathscr{N}_{2}^{SL}(\tau)=\left\{\begin{array}{llc}
A & \mathrm{for} & -\frac{T}{2}\leq\tau<\tau'_{m} \\
\frac{2}{\pi}\frac{\langle N_{3}^{(1,3)}\rangle_{\mathrm{eff}}}{\tilde{s}_{0}\langle N_{3}^{(1,3)}\rangle_{\mathrm{eff}}^{1/3}}\cos^{2}{\left(\frac{\tau-\tau_{s}}{\tilde{s}_{0}\langle N_{3}^{(1,3)}\rangle_{\mathrm{eff}}^{1/3}}\right)} & \mathrm{for} & \tau'_{m}\leq\tau\leq\tau''_{m} \\
B & \mathrm{for} & \tau''_{m}<\tau\leq\frac{T}{2} \end{array}\right.
\end{equation}
for the values
\begin{subequations}
\begin{eqnarray}
\tau'_{m}&=&\tau_{s}-\tilde{s}_{0}\langle N_{3}^{(1,3)}\rangle_{\mathrm{eff}}^{1/3}\cos^{-1}{\sqrt{\frac{\pi A\tilde{s}_{0}}{2\langle N_{3}^{(1,3)}\rangle_{\mathrm{eff}}^{2/3}}}}\\
\tau''_{m}&=&\tau_{s}+\tilde{s}_{0}\langle N_{3}^{(1,3)}\rangle_{\mathrm{eff}}^{1/3}\cos^{-1}{\sqrt{\frac{\pi B\tilde{s}_{0}}{2\langle N_{3}^{(1,3)}\rangle_{\mathrm{eff}}^{2/3}}}}
\end{eqnarray}
\end{subequations}
of the discrete time coordinate that match the discrete spatial $2$-volumes $A$ and $B$ of the stalks to those of the central accumulation. For minimal boundary--nonminimal boundary transition amplitudes when stalks are not present, we employ the function \eqref{discretedSvolprofileminnonmin} with $\tau'_{m}=-\frac{T}{2}$ and $\tau''_{m}=\frac{T}{2}$, and we enforce that the function \eqref{discretedSvolprofileminnonmin} passes through the boundary values of $\langle N_{2}^{SL}(\tau)\rangle$. The values of $\tilde{s}_{0}$ and $\tau_{s}$ are thus determined to be the roots of the transcendental equations
\begin{subequations}
\begin{eqnarray}
4&=&\frac{2}{\pi}\frac{\langle N_{3}^{(1,3)}\rangle_{\mathrm{eff}}}{\tilde{s}_{0}\langle N_{3}^{(1,3)}\rangle_{\mathrm{eff}}^{1/3}}\cos^{2}{\left(\frac{\frac{T}{2}+\tau_{s}}{\tilde{s}_{0}\langle N_{3}^{(1,3)}\rangle_{\mathrm{eff}}^{1/3}}\right)}\\
X&=&\frac{2}{\pi}\frac{\langle N_{3}^{(1,3)}\rangle_{\mathrm{eff}}}{\tilde{s}_{0}\langle N_{3}^{(1,3)}\rangle_{\mathrm{eff}}^{1/3}}\cos^{2}{\left(\frac{\frac{T}{2}-\tau_{s}}{\tilde{s}_{0}\langle N_{3}^{(1,3)}\rangle_{\mathrm{eff}}^{1/3}}\right)}
\end{eqnarray}
\end{subequations}
since $\langle N_{2}^{SL}(\mathcal{S}_{i}^{2})\rangle=4$ and $\langle N_{2}^{SL}(\mathcal{S}_{f}^{2})\rangle=X$ in this case.

For nonminimal boundary--nonminimal boundary transition amplitudes when stalks are present, we employ in particular the function 
\begin{equation}\label{variantdiscretedSvolprofileappen1}
\mathscr{N}_{2}^{SL}(\tau)=\left\{\begin{array}{llc}
A & \mathrm{for} & -\frac{T}{2}\leq\tau<\tau'_{m} \\
\frac{2\langle N_{3}^{(1,3)}\rangle\cos^{2}{\left(\frac{\tau-\tau_{s}}{\tilde{s}_{0}\langle N_{3}^{(1,3)}\rangle^{1/3}}\right)}}{\tau_{f}-\tau_{i}+\tilde{s}_{0}\langle N_{3}^{(1,3)}\rangle^{1/3}\sin{\left(\frac{\tau_{f}-\tau_{i}}{\tilde{s}_{0}\langle N_{3}^{(1,3)}\rangle^{1/3}}\right)}\cos{\left(\frac{\tau_{f}+\tau_{i}}{\tilde{s}_{0}\langle N_{3}^{(1,3)}\rangle^{1/3}}\right)}} & \mathrm{for} & \tau'_{m}\leq\tau\leq\tau''_{m} \\
B & \mathrm{for} & \tau''_{m}<\tau\leq\frac{T}{2} \end{array}\right.
\end{equation}
for the values
\begin{subequations}
\begin{eqnarray}
\tau'_{m}&=&\tau_{s}-\tilde{s}_{0}\langle N_{3}^{(1,3)}\rangle^{1/3}\cos^{-1}{\sqrt{\frac{A\left[\tau_{f}-\tau_{i}+\tilde{s}_{0}\langle N_{3}^{(1,3)}\rangle^{1/3}\sin{\left(\frac{\tau_{f}-\tau_{i}}{\tilde{s}_{0}\langle N_{3}^{(1,3)}\rangle^{1/3}}\right)}\cos{\left(\frac{\tau_{f}+\tau_{i}}{\tilde{s}_{0}\langle N_{3}^{(1,3)}\rangle^{1/3}}\right)}\right]}{2N_{3}^{(1,3)}}}}\nonumber\\
\\
\tau''_{m}&=&\tau_{s}+\tilde{s}_{0}\langle N_{3}^{(1,3)}\rangle^{1/3}\cos^{-1}{\sqrt{\frac{B\left[\tau_{f}-\tau_{i}+\tilde{s}_{0}\langle N_{3}^{(1,3)}\rangle^{1/3}\sin{\left(\frac{\tau_{f}-\tau_{i}}{\tilde{s}_{0}\langle N_{3}^{(1,3)}\rangle^{1/3}}\right)}\cos{\left(\frac{\tau_{f}+\tau_{i}}{\tilde{s}_{0}\langle N_{3}^{(1,3)}\rangle^{1/3}}\right)}\right]}{2N_{3}^{(1,3)}}}}\nonumber\\
\end{eqnarray}
\end{subequations}
of the discrete time coordinate that match the discrete spatial $2$-volumes $A$ and $B$ of the stalks to those of the central accumulation. For nonminimal boundary--nonminimal boundary transition amplitudes when stalks are not present and $N_{2}^{SL}(\mathcal{S}_{i}^{2})=N_{2}^{SL}(\mathcal{S}_{f}^{2})$, we employ the function \eqref{variantdiscretedSvolprofileappen} with $\tau'_{m}=-\frac{T}{2}$ and $\tau''_{m}=\frac{T}{2}$, and we enforce that the function \eqref{variantdiscretedSvolprofileappen1} passes through the boundary values of $\langle N_{2}^{SL}(\tau)\rangle$. The value of $\tilde{s}_{0}$ is thus determined to be the root of the transcendental equation 
\begin{equation}
X=\frac{2\langle N_{3}^{(1,3)}\rangle\cos^{2}{\left(\frac{\frac{T}{2}}{\tilde{s}_{0}\langle N_{3}^{(1,3)}\rangle^{1/3}}\right)}}{\tau_{f}-\tau_{i}+\tilde{s}_{0}\langle N_{3}^{(1,3)}\rangle^{1/3}\sin{\left(\frac{\tau_{f}-\tau_{i}}{\tilde{s}_{0}\langle N_{3}^{(1,3)}\rangle^{1/3}}\right)}\cos{\left(\frac{\tau_{f}+\tau_{i}}{\tilde{s}_{0}\langle N_{3}^{(1,3)}\rangle^{1/3}}\right)}}
\end{equation}
since $N_{2}^{SL}(\mathcal{S}_{i}^{2})=N_{2}^{SL}(\mathcal{S}_{f}^{2})=X$ in this case.

To determine the best fit of the function \eqref{discretedSvolprofileapp} to $\langle N_{2}^{SL}(\tau)\rangle$, we minimize the chi-squared function
\begin{equation}\label{chisquared}
\chi^{2}(\tilde{s}_{0},\ldots)=\sum_{\tau=1}^{T}\frac{\left[\langle N_{2}^{SL}(\tau)\rangle-\mathscr{N}_{2}^{SL}(\tau)\right]^{2}}{\sigma^{2}(\langle N_{2}^{SL}(\tau)\rangle)}
\end{equation}
with the ellipsis indicating the potential inclusion of the fit parameters $A$, $B$, and $\tau_{s}$. Here,
\begin{equation}
\sigma^{2}(\langle N_{2}^{SL}(\tau)\rangle)=\frac{\langle(N_{2}^{SL}(\tau))^{2}\rangle-\langle N_{2}^{SL}(\tau)\rangle^{2}}{\sqrt{N(\mathcal{T}_{c})}}
\end{equation}
is the measured variance in $\langle N_{2}^{SL}(\tau)\rangle$. We determine the error $\epsilon(\tilde{s}_{0})$ in the fit parameter $\tilde{s}_{0}$ by solving for the two values of $\tilde{s}_{0}$ satisfying
\begin{equation}
\chi^{2}(\tilde{s}_{0},\ldots)=\chi^{2}_{\mathrm{min}}(\tilde{s}_{0},\ldots)+1,
\end{equation}
one less than and one greater than the value of $\tilde{s}_{0}$ at the minimum of the chi-squared function \eqref{chisquared}. We set the fit parameters $A$, $B$, and $\tau_{s}$ to their values at the minimum of the chi-squared function \eqref{chisquared} in making this determination. Our measure of the goodness of fit of the function \eqref{discretedSvolprofileapp} to $\langle N_{2}^{SL}(\tau)\rangle$ is the minimum chi-squared per degree of freedom
\begin{equation}
\chi^{2}_{\mathrm{pdf}}=\frac{\chi^{2}_{\mathrm{min}}(\tilde{s}_{0},\ldots)}{T-n},
\end{equation}
where $n$ is the number of fit parameters.

\end{document}